\pdfoutput=1
\documentclass[twocolumn]{aastex63}

\newcommand{\ts}{\textsuperscript}

\shorttitle{Ne V-strong PSB QSO}
\shortauthors{Swiggum et al.}

\usepackage{listings}
\usepackage{color}
\usepackage{float}
\usepackage{tabularx}
\usepackage{hyphenat}

\definecolor{dkgreen}{rgb}{0,0.6,0}
\definecolor{gray}{rgb}{0.5,0.5,0.5}
\definecolor{mauve}{rgb}{0.58,0,0.82}

\begin{document}
\title{Understanding the Nature of an Unusual Post-Starburst Quasar with Exceptionally
Strong Ne V Emission}

\correspondingauthor{Cameren Swiggum}
\email{cameren.swiggum@univie.ac.at}
\author[0000-0001-9201-5995]{Cameren Swiggum}\affiliation{Department of Astronomy, University of Wisconsin-Madison, Madison, WI 53706, USA}

\author[0000-0003-3097-5178]{Christy Tremonti}
\affil{Department of Astronomy, University of Wisconsin-Madison, Madison, WI 53706, USA}

\author{Serena Perrotta}
\affil{Center for Astrophysics and Space Sciences, University of California, San Diego, La Jolla, CA 92093, USA}

\author{Adam Schaefer}\affil{Department of Astronomy, University of Wisconsin-Madison, Madison, WI 53706, USA}
\affil{Max Planck Institute for Astrophysics
Karl-Schwarzschild-Strasse 1, 85748 Garching, Germany}

\author[0000-0003-1468-9526]{Ryan C. Hickox}
\affil{Department of Physics and Astronomy, Dartmouth College, Hanover, NH 03755, USA}

\author[0000-0002-2583-5894]{Alison L. Coil}
\affil{Center for Astrophysics and Space Sciences, University of California, San Diego, La Jolla, CA 92093, USA}

\author[0000-0003-1771-5531]{Paul H. Sell}
\affil{Department of Astronomy, University of Florida, Gainesville, FL, 32611 USA}

\author{Aleksandar M. Diamond-Stanic}
\affil{Department of Physics and Astronomy, Bates College, Lewiston, ME, 04240, USA}

\author{Jalyn Krause}
\affil{Department of Astronomy, University of Wisconsin-Madison, Madison, WI 53706, USA}

\author{Gregory Mosby}
\affil{NASA Goddard Space Flight Center, Mail Code:665,
Greenbelt, MD 20771}

\begin{abstract}

We present a $z = 0.94$ quasar, SDSS J004846.45-004611.9, discovered in the SDSS-III BOSS survey. A visual analysis of this spectrum reveals highly broadened and blueshifted narrow emission lines, in particular [Ne~V]$\lambda3426$ and [O~III]$\lambda5007$, with outflow velocities of 4000 km s$^{-1}$, along with unusually large [Ne V]$\lambda3426$/[Ne III]$\lambda3869$ ratios. The gas shows higher ionization at higher outflow velocities, indicating a connection between the powerful outflow and the unusual strength of the high ionization lines. The SED and the $i - \text{W3}$ color of the source reveal that it is likely a ``core'' Extremely Red Quasar (core ERQ); a candidate population of young AGN that are violently ``blowing out'' gas and dust from their centers. The dominance of host galaxy light in its spectrum and its fortuitous position in the SDSS S82 region allows us to measure its star formation history and investigate for variability for the first time in an ERQ. Our analysis indicates that SDSS J004846.45-004611.9 underwent a short-lived starburst phase 400 Myr ago and was subsequently quenched, possibly indicating a time-lag between star formation quenching and the onset of AGN activity. We also find that the strong extinction can be uniquely attributed to the AGN and does not persist in the host galaxy, contradicting a scenario where the source has recently transitioned from being a dusty sub-mm galaxy. In our relatively shallow photometric data, the source does not appear to be variable at $0.24-2.4~\mu$m in the restframe, most likely due to the dominant contribution of host galaxy starlight at these wavelengths.

\end{abstract}




\section{Introduction} \label{introduction}
Quasars are galaxies host to active galactic nuclei (AGN): central regions of galaxies where rapid accretion of matter onto a supermassive black hole (SMBH) occurs. These AGN drive some of the most powerful outflows of gas and dust observed in the Universe. An outstanding problem in galaxy formation and evolution concerns the relationship between these AGN-driven outflows and the resultant effects on their host galaxies. Observational studies have revealed quasar outflows as a potential source of star formation regulation and/or quenching, as feedback from AGN may violently disrupt the surrounding interstellar medium (ISM) of their host galaxies \citep[e.g.][]{Blandford_2004, Scannapieco_2004, Vernaleo_2006, Fabian_2012, Kormendy_Ho_2013}.

Theoretical models of galaxy evolution indicate that quasar feedback may result from major merger events that trigger rapid gas inflows \citep[e.g.][]{Hopkins_2005, Hopkins_2006, Hopkins_2008, Blecha_2018}. These events lead to a burst in star formation along with rapid accretion onto the central SMBH. The dramatic disruption of dust and gas due to these mergers causes galaxies at this stage of evolution to appear observationally as sub-mm galaxies or ultra-luminous infrared galaxies \citep[e.g.][]{Sanders_1988, Veilleux_2009, Simpson_2014}. Subsequently, energy and momentum from the accreting SMBH couples to the surrounding ISM, causing a galactic-scale ``blowout'' of gas and dust, thus revealing a visibly luminous quasar \citep[e.g.][]{Sanders_1988, Di_matteo_2005, Hopkins_2016, Rupke_2011, Rupke_2013, Liu_2013}.

Observing populations of dust-reddened quasars \citep{Urrutia_2008, Banerji_2012, Glikman_2012} in the process of blowing out gas and dust from their central SMBHs can give insight into the immediate effects AGN have on their host galaxies. Observationally, this task is difficult as the blowout phase is estimated to be relatively brief on the time scale of galaxy evolution. Wide field spectroscopic surveys such as Sloan Digital Sky Survey III \citep[SDSS-III;][]{Eisenstein_2011} Baryon acoustic Oscillations Spectroscopic Survey \citep[BOSS;][]{Dawson_2013} and their vast data sets yield the ability to compile statistical samples of galaxies at many evolutionary stages. 
Recently, a population of extremely red quasars \citep[ERQs;][]{Ross_2015, Hamann_2017} with a redshift peak around $z \sim 1$ and a second peak at $z \sim 2 - 3$ were discovered in the SDSS Data Release 12 (DR12). These quasars are selected by their rest frame UV to mid-IR colors from SDSS and the Wide-field Infrared Survey Explorer \citep[WISE;][]{Wright_2010, Lang_2016} as $i - \text{W3} > 4.6$ (AB). BOSS spectra reveal they are host to a suite of peculiar emission features in the rest frame UV, of which most notable is the broad, blueshifted and wingless emission of C IV $\lambda1549$. \cite{Hamann_2017} imposes an additional selection of $\text{REW(C IV)} > 100 \text{\AA}$ to define a sub-sample of ``core ERQs'' which have a more homogeneous suite of emission line properties and represent the strongest candidates for AGN in the blowout phase. The distinctive spectral features of core ERQs insure that they are not simply Type I quasars viewed behind an unusually large dust reddening screen.
Additionally, their overall spectral shape from UV to far-IR wavelengths is intrinsically different from Type 1 and Type 2 quasars, as well as from a Type 1 quasars behind a dust-reddening screen \citep[cf.][Figure 16]{Hamann_2017}. 

Extremely broad and blueshifted [O~III]~$\lambda\lambda4959,5007$ has been observed in the near-infrared (NIR) spectra of 24 core ERQs by \cite{Perrotta_2019}. [O~III] is found to be a suitable tracer for \emph{galactic-scale}
outflows as the line is a forbidden transition which cannot be produced in the high density, sub-parsec scales of the broad-line region (BLR) near the central AGN. Instead, it is abundant at the low density kiloparsec scales of the narrow-line region (NLR). The kinematics of the [O~III] line have been used to study galactic-scale outflows in Type II AGN over a wide range in luminosity and redshift \citep[e.g.][]{Mullaney_2013, Zakamska_2014, Harrison_2014, Carniani_2015, Carniani_2016}. The core ERQs of \cite{Perrotta_2019} have [O~III] widths ($w_{90}$) in the range of 2053 - 7227 km s$^{-1}$. For comparison, a sample of typical blue quasars from \cite{Shen_2016} with bolometric luminosities in the range of ERQs ($10^{46} - 10^{48}$ ergs s$^{-1}$) have a $w_{90}$ range of $\sim$ 1000 - 2000 km s$^{-1}$ \citep[cf.][Figure 6]{Perrotta_2019}. 

Additionally, \cite{Perrotta_2019} found the core ERQs to exhibit a strong correlation between [O~III] outflow strength and $i-\text{W3}$ color, suggesting that the dust surrounding the AGN may assist in the coupling of energy and momentum between the AGN and the ISM. This relationship has been found in other Type 1 and 2 quasar samples as well \citep{DiPompeo_2018}. This process may stem from optical and UV radiation emitted by the central AGN being absorbed by the surrounding dust and re-emitted at IR wavelengths. In the case that the dusty envelope has a high IR optical depth ($\tau_{\text{IR}} > 1$) and sufficiently encompasses the central AGN, the IR photons will repeatedly scatter, building up momentum which transfers into the surrounding ISM. Analytical calculations \citep[e.g.][]{Murray_2005, Thompson_2015, Ishibashi_2015, Ishibashi_2016} and radiative transfer calculations \citep[e.g.][]{Proga_2004, Krumholz_2012, Krumhoz_2013, Bieri_2017} have noted that this coupling mechanism can accelerate AGN winds. 


Here we report on SDSS J004846.45-004611.9 (hereafter J0048-0046), a $z\sim1$ quasar with a dominant host galaxy continuum in addition to broad, blueshifted high ionization emission lines of which most notable is the forbidden [Ne~V]$\lambda3426$ line. Its spectrum was identified serendipitously via visual inspection of a subset of spectra from BOSS. Upon calculation of its optical-to-IR color, we find that it meets the ERQ selection criterion of $i-\text{W3} > 4.6$. The goal of our study is to determine if J0048-0046 is a candidate \textit{young} AGN driving powerful outflows and whether its novel observational properties such as its spectroscopically dominant host galaxy and its available time-series photometry can yield new insight into the physical mechanisms of these sources. 

This paper is organized as follows: In Section \ref{sec:data}, we lay out the multitude of spectroscopic and photometric datasets used to investigate the multi-wavelength and time-domain properties of J0048-0046. In Section \ref{sec:spectral_analysis}, we analyze the multi-wavelength spectral properties of J0048-0046 and report on its emission line profiles. In Section \ref{sec:host_galaxy_analysis}, we analyze the morphology and stellar continuum to infer its star formation history. In Section \ref{sec:variability}, we probe optical and IR variability of J0048-0046 at day-to-month-long timescales. We interpret our results in Section \ref{sec:results} and make comparisons to other studies of ERQs and quasars in general. We conclude with a summary of the work in Section \ref{sec:summary}. Throughout our analysis, we adopt a $\Lambda$CDM cosmology of $\Omega_{M} = 0.315$, $\Omega_{\Lambda} = 0.685$, and H$_{0}$ $= 67.4$ km s$^{-1}$ Mpc$^{-1}$ from the \cite{Planck_2020}.  All derived stellar masses and star formation rates (SFRs) assume a \cite{Salpeter_1955} initial mass function (IMF).

\section{Data}
\label{sec:data}

Quasars exhibit a multitude of spectral properties over a wide range of wavelengths. To gain a better understanding of the physical processes at work in J0048-0046, we assemble data from a variety of ground and space-based telescopes tracing the spectral regime from far-UV to radio wavelengths.

\begin{figure*}[!ht]
    \centering
    \includegraphics[width = \textwidth]{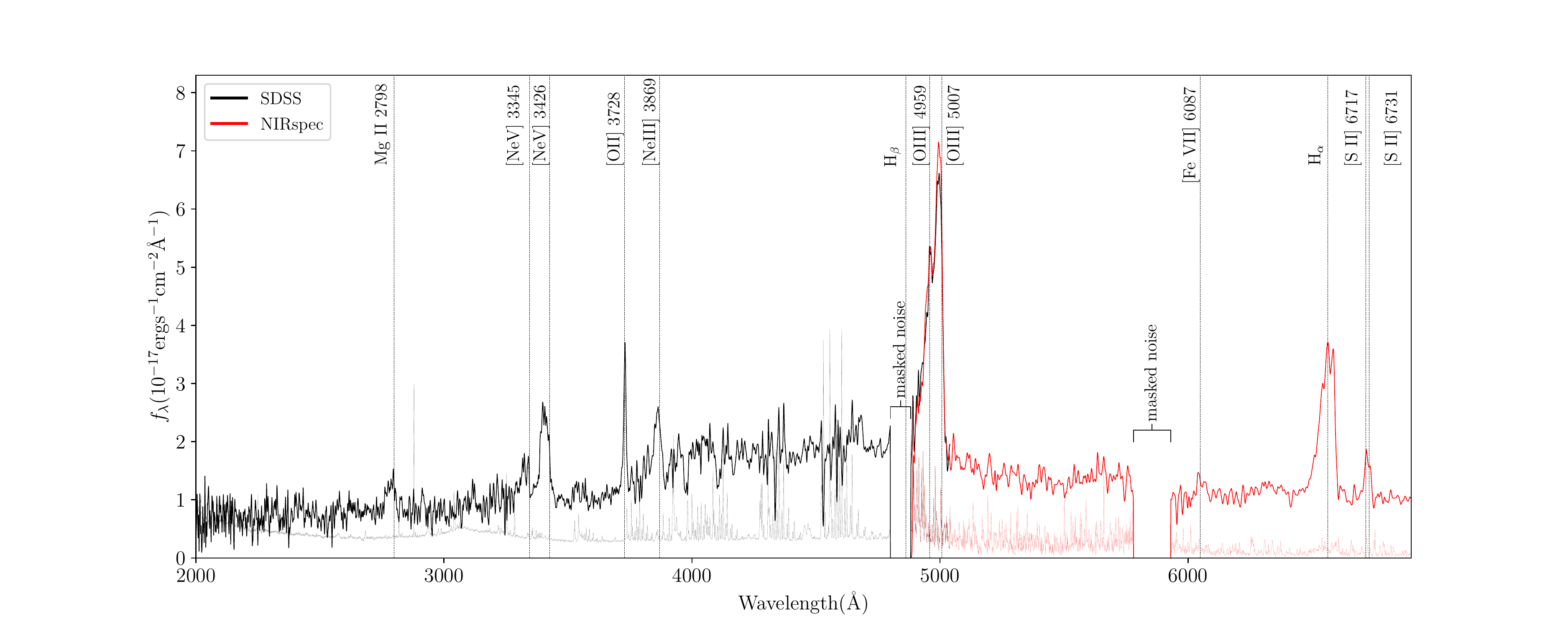}
    \caption{Combined restframe SDSS BOSS (black) and NIRSPEC (red) spectra of J0048-0046. Prominent emission lines are labeled across the top. The error spectrum for both BOSS and NIRspec and runs across the bottom, peaking at regions most affected by sky-subtraction residuals. The wavelength region between 5780 \AA\ and 5930 \AA\ is masked due to poorly subtracted sky lines.}
    \label{spectrum}

\end{figure*}

\subsection{Optical and Near Infrared Spectra}
 
 \subsubsection{SDSS-III/BOSS}
 J0048-0046 was selected for spectroscopic follow-up in SDSS-III \citep{Eisenstein_2011} as part of the `Reddened Quasars' ancillary program 
 \footnote{\url{https://www.sdss.org/dr16/algorithms/ancillary/boss/redqso/}} which attempted to target quasars with E(B-V) $>$ 0.5 in the Stripe 82 area using a combination of SDSS, 2MASS, and FIRST photometry. 
 SDSS-III uses the optical 2.5 meter wide-angle telescope at Apache Point Observatory \citep{Gunn_2006}. The BOSS spectrograph \citep{Smee_2013} uses fiber-based spectroscopy with 2" diameter optical fibers. Light that traverses through the fiber is beam-split into a blue channel and a red channel with corresponding wavelength ranges of $3600-6350$ \AA\ and $5650-10,000$~\AA. The spectral resolution is in the range of $R = 1560$ at 3700 $\text{\AA}$ and $R = 2650$ at 6000 $\text{\AA}$. Reduction procedures are carried out by the BOSS data reduction pipeline \textit{idlspec2d} described in \cite{Dawson_2013}. 
 
 The BOSS spectrum for J0048-0046 was taken on January 5\ts{th}, 2010 and is a combination of eight individual exposures, each with fifteen minute integration time. The reported SDSS pipeline redshift is $z = 6.937$, however, the spectral fit mistook a broad [OIII] 5007$\text{\AA}$ line for Ly$\alpha$. We instead use the reported redshift value \verb|z_noqso| $= 0.9379 \pm 0.0001$ which omits QSO templates when determining the best-fit spectral template. The SDSS  spectrum (see Figure \ref{spectrum}) covers $2500-5050$ \AA\ in the quasar's rest frame.

\subsubsection{SALT/RSS}

In order to more accurately constrain the kinematics of the [Ne~V] $\lambda3426$ line, we obtained follow up spectrum with higher S/N around [Ne~V] using the South African Large Telescope (SALT) Robert Stobie Spectrograph \citep[RSS; ][]{Kobulnicky_2003} (2015-1-SCI-020, P.I.: C. Tremonti). The observation was carried out in long-slit mode on August 8\ts{th}, 2015 with seeings of 1.3" - 1.5". We used the RSS PG1300 grating with a 28 deg tilt angle, which provides a resolving power of $R\sim2200$ for our observed wavelength range of 6325 \AA{} and 8148 \AA{} ($3260 - 4200 \text{\AA}$ in the target's restframe). Six exposures of J0048-0046 were taken with the RSS at 1140 seconds each. Gain-correction, bias-correction, and mosaicking of these exposures were performed by the SALT data reduction pipeline \citep[][]{Crawford_2010} and further reduced using PyRAF. Wavelength calibration solutions as a function of pixel positions were found by identifying lines of a Xenon arc lamp. These solutions were subsequently applied to each exposure. Cosmic ray removal and background subtraction were performed for each exposure. They were then median combined into one 2D spectrum. We then applied extinction correction, aperture extraction, and a heliocentric correction to obtain our final 1D spectrum.

\subsubsection{Keck/NIRSPEC} 

Spectra were obtained with the Near Infrared Spectrometer (NIRSPEC; McLean et al. 1998) on the Keck II telescope on September 15th and 17th, 2013. We used the NIRSPEC filters 1 and 3 corresponding to the Y and J bands to obtain spectral coverage from 0.95 - 1.35 $\mu$m which includes the [O~III] $\lambda\lambda5007,4959$ and H$\alpha$ lines in the quasar's restframe.  We used the $0\farcs76 \times 42$ arcsecond longslit for a spectral resolution of $R=2000$ or 150 km~s$^{-1}$. A series of $8\times300$ seconds in the NIRSPEC-1 filter and $12\times300$ exposures in the NIRSPEC-2 filter were taken at two different positions along the slit, employing the standared ABBA slit nodding pattern. 

The REDSPEC \footnote {\url{https://www2.keck.hawaii.edu/inst/nirspec/redspec.html}} IDL package was used to make spatial and spectral corrections to the raw data.  The exposures were dark subtracted and flat-fielded using an internal
flat-field calibration lamp.
Sky subtraction was preformed by subtracting A-B pairs. We performed relative flux calibrations and telluric absorption corrections using spectra of standard stars observed the same night.  The wavelengths are calibrated to the vacuum heliocentric system using spectra of internal arc lamps.  Following standard pipeline produces, we extracted calibrated spectra which were averaged to produce a final spectrum. The standard deviation was used to create a final error spectrum.  
 Absolute flux calibration was not possible due to variable cloud cover. We therefore scale the NIRSPEC spectrum to match the SDSS spectrum in the 5030 - 5290 \AA\ region in the restframe which is free of emission lines.  The SDSS and NIRSPEC data both cover the [OIII] $\lambda\lambda$~4959,5007 emission lines; the agreement between the two spectra is excellent (see Fig.~\ref{spectrum}). Analysis of the co-added 2-dimensional spectrum does not reveal any spatially extended [O~III]~$\lambda\lambda 4959,5007$.


\subsection{SDSS and WISE photometry} \label{sec:photo_data}
 J0048-0046 is positioned within the SDSS Stripe 82 region (S82), a~300 sq deg region on the Celestial Equator that was repeatedly imaged in the \textit{ugriz} filters to create a unique database for studying variability \citep[c.f.][]{Abazajian_2009}. We use eleven epochs of the photometric data products output from the official SDSS data release images, with the latest being from Data Release 16 (DR16). Hereafter, we refer to all eleven epochs as DR16. These data are reduced using the SDSS photometric pipeline \citep{Lupton_2001, Stoughton_2002} and the product is located on the Catalog Archive Server (CAS\footnote{\url{http://skyserver.sdss.org/CasJobs/SubmitJob.aspx}}), which provides a Structured Query Language (SQL) interface for obtaining product data. 

To extract morphological parameters and constrain parameters for proper variability analysis, we utilize the deep co-add images of the S82 database from \cite{Fliri_2016}, which reach $3\sigma$ values of 24.2, 25.2, 24.7, 24.3, 23.0 mag in the \textit{ugriz} filters. Since the S82 co-adds are on average four magnitudes deeper than individual SDSS images, we apply our own photometric analysis to these co-adds of J0048-0046 and its neighboring galaxy (Figure \ref{nev_image}) using the GALFITM \citep{Vika_2013}, a photometric-fitting code with its core procedures inherited from the GALFIT package \cite{Peng_2002}. GALFITM incorporates the ability to fit morphological parameters as a function of wavelength, which is useful for our available multi-band images.

Our analysis provides us with an estimate of the effective radius ($R_{50}$) of J0048-0046 to be compared with the sub-sample of HST-imaged ERQs from \cite{Zakamska_2019}. PSF images for the field are provided by \cite{Fliri_2016}, which are input to GALFITM and are incorporated by the program before the model fitting is performed. Sigma images are created in-house by GALFITM through retrieval of the FITS header keys EXPTIME, GAIN, and NCOMBINE (set to 303). We fit J0048-0046 with both a PSF and a sersic model simultaneously, whereas we fit the comapion galaxy with a sersic model only. Morphological fitting parameters, including the sources' effective radii (R50), sersic indices, axis ratios, and position angles are fit as a quadratic function of wavelength, while source sersic/psf magnitudes and centroids are allowed to freely vary. Initial values are informed from the SDSS DR16 photObj table. GALFITM performs the fitting on both J0048 and the companion simultaneously using the Levenberg–Marquardt minimization algorithm.

Since the companion galaxy's centroid is only $\sim2$ arcseconds away from that of J0048-0046, it is possible that nightly changes in seeing affect the SDSS pipeline's deblending of the two sources; flux measurements of J0048-0046 could see some epochs including contaminating light from the companion. We perform additional photometric analysis on the 11 SDSS data release images of J0048-0046 and the companion in each of the \textit{griz} bands. We follow the same method using GALFITM as applied to the S82 co-add, however we only fit a sersic model to J0048-0046 and neglect the PSF model as the low S/N in some of the images causes issues with parameter degeneracy (specifically in the magnitudes and centroids) when using both sersic and PSF models. The starting parameters for the sersic models of J0048-0046 and its compainon are informed by the S82 co-add results. For each epoch, we follow the prescription of \cite{Anderson_2000} and its Python implementation from \cite{Bradley_2021} to construct PSF images from high S/N stars in the same field as J0048-0046. The best-fit GALFITM model parameters for J0048-0046 and its companion are quoted in Appendix \label{galfit}.
 
 The Wide-field Infrared Survey Explorer \citep[WISE; ][]{Wright_2010} provides infrared magnitudes in filters centered at 3.4, 4.6, 12 and 22 $\mu$m, (hereafter W1, W2, W3, W4, respectively.) 
 These data are used to construct J0048-0046's Spectral Energy Distribution (SED, \S\ref{sec:sed_analysis}) and to calculate its bolometric luminosity (Table \ref{lum_table}). We utilize photometry from the unWISE co-adds \citep{Lang_2014} which provide better resolution and S/N than the original co-add images.  We note that the minor axis FWHMs of the four filters range from 5.6'' to 11.65'' meaning that J0048-0046 and its neighboring galaxy $\sim$2'' away (Figure \ref{nev_image}) are not resolved separately.
 However, in \S\ref{sec:sed_analysis}, we show that the bluer and less luminous companion (in the \textit{u, r} and \textit{z} bands) galaxy is unlikely to be contributing more than $\sim5$\% of the flux in the WISE bands. WISE luminosities were calculated by first converting the reported Vega magnitudes to flux densities and using the appropriate color corrections \citep{Wright_2010}. The WISE W1-W4 data reported in Table \ref{lum_table} were obtained in the middle of 2010. Multi-year detections in the W1 and W2 bands are also available from late-2013 until late-2018 through the NEOWISE survey \citep[][]{Mainzer_2011}, providing a time-series dataset that we use to investigate IR variability in J0048-0046.

\subsection{Other Photometric data}
Other sources of photometric data were obtained to further study the presence of optical variability in J0048-0046 and to complete its SED.

The Panoramic Survey Telescope and Rapid Response System \citep[PanSTARRS1-PS1; ][]{Chambers_2016} data release 2 extends our catalog of optical photometry for our SED in the \textit{grizy} passbands. Data was obtained using the MAST CAS SQL\footnote{\url{https://catalogs.mast.stsci.edu}} interface, extracting dates, Kron magnitudes, their respective errors, and photometric flags. While multiple epochs of data are taken of J0048-0046, they are sparse and do not meet our requirements for accurate variability analysis. 



For the purpose of classifying J0048-0046 as either radio-quiet (RQ) or radio-loud (RL) and in order to draw connection to ionized outflows, we use a radio flux density measurement at 1435 MHz provided by the Faint Images of the Radio Sky at Twenty Centimeters \citep[FIRST; ][]{Becker_1995} survey which utilized the NRAO Very Large Array. 

Additionally, we construct a more complete baseline for our SED by utilizing photometric data in the UV and near-IR. NASA's Galaxy Evolution Explorer \citep[GALEX;][]{Martin_2005} provides one epoch of ultraviolet data in the far-UV ($\lambda = 1258 \text{\AA}$) and the near-UV ($\lambda = 2310 \text{ \AA}$). The images have FWHMs of 4.2''and 5.3'' in the far- and near-UV respectively, and therefore, light from the companion galaxy 2'' away will contribute a fraction of the observed flux. We consider this further in \S\ref{sec:sed_analysis}.

The United Kingdom Infrared Telescope Deep Sky Survey \citep[UKIDSS;][]{Lawrence_2007}  provides fluxes from its Wide Field Camera \citep[WFCAM;][]{Casali_2007} in the \textit{Y} ($\lambda = 1.03 \: \mu \text{m}$), \textit{J} ($\lambda = 1.25 \: \mu \text{m}$), \textit{H} ($\lambda = 1.63 \: \mu \text{m}$), and \textit{K} ($\lambda = 2.20 \: \mu \text{m}$) filters.  Measurements of total magnitudes for extended sources are problematic in the UKIDSS public catalogs.  We therefore use the APER3 magnitudes corrected following the procedure outlined in \citet{Bundy_2015}.

To display a color image of the field surrounding J0048-0046 (Figure \ref{color_image}), we use publicly available data from the Dark Energy Camera Legacy Survey (DECaLS) of the SDSS Equatorial Sky Data Release 8 (PI: D. Schlegel and A. Dey), observed by the Dark Energy Camera \citep[DECam;][]{Flaugher_2015} and part of the 14,000 $\text{deg}^{2}$ DESI Legacy Imaging Surveys \citep{Dey_2019}. Data consist of observations in the \textit{grz} optical bands with depths of at least $g=24.0, \; r=23.4, \text{ and } z=22.5$ magnitude, $1 - 2$ magnitude deeper than SDSS photometry.

\subsection{Bolometric and Radio Luminosity}  
\label{sec:luminosities}
Calculating the bolometric and radio luminosity of J0048-0046 is necessary to get a complete view of the radiative and mechanical power of this source, and understand how this source compares to the full population of quasars as specified by the quasar luminosity function. We use a combination of infrared and optical data to estimate the bolometric luminosity of J0048-0046 and the FIRST radio flux density for the radio luminosity. Our adopted cosmological parameters (Section \ref{introduction}) with the estimated redshift are used to infer the luminosity distance to J0048-0046.

The infrared flux values from the WISE W3 and W4 passbands are used to infer AGN bolometric luminosities via the bolometric corrections defined in \cite{Runnoe_2012}. We follow a procedure for obtaining the W3 and W4 luminosities by first converting the reported Vega magnitudes to flux densities (Jy). We follow flux conversion methods and color corrections\footnote{\url{http://wise2.ipac.caltech.edu/docs/release/allsky/expsup/sec4_4h.html}} with guidance from \cite{Wright_2010}. We use the conversion given as 

$$ F_{\nu}[Jy] = \frac{F^*_{\nu 0}}{f_c} \times 10^{-m_{vega}/2.5}$$

\noindent where $F^*_{\nu 0}$ is the zero magnitude flux density for sources following a power law spectrum of $F_\nu \propto \nu^{-(2)}$ and $f_c$ is the color correction factor for sources deviating from such a power law ($F_\nu \propto \nu^{-(\alpha)}$). According to our reported W2-W3 and W3-W4 colors, the mid-IR spectra of J0048-0046 is approximately a power law with $\alpha = 2$, hence $f_c = 1$. Additionally, it is reported that a correction to the W4 flux density must be made for steeply rising mid-IR spectra ($\alpha \geq 1$) due to calibration uncertainties. This correction follows as: $F'_{\nu}\text{[W4]} \approx 0.90 \times F_{\nu}\text{[W4]}$. We use equations 6 and 8 from \cite{Runnoe_2012} to calculate our source's bolometric luminosity from its W3 and W4 luminosities. 

Estimates on AGN bolometric luminosity are also calculated using the [O~III]~$\lambda5007$ and [Ne~V]~$\lambda3426$ emission lines. We calculate the luminosity of both lines using the combined narrow and broad components of the fits described in §3.1. We adopt equation 3 from \cite{Pennell_2017} to compute an [O~III]$\lambda5007$ bolometric luminosity. Since [Ne~V]$\lambda3426$ has a much higher ionization potential than [O~III], it is an even better tracer of AGN activity. However, there are no published bolometric corrections for [Ne~V]. We instead use the work of \cite{Gilli_2010}, who computed the relationship between $2 - 10$ keV (X-ray) and [Ne~V]$\lambda3426$ using a sample of  AGN at $z = 0 - 1.5$, finding $L_{X}/L_{Ne~V} \sim 400.$  We then utilize luminosity-dependent X-ray bolometric correction of \citet[][their equation 3]{Duras_2020}. Taking the median luminosity of the Gilli sample to be  $L_{2-10~\text{eV}} = 44.5$, we find an X-ray bolometric correction of $K_X$ = 22.12.   Combining this with the $L_{X}/L_{Ne~V}$ ratio, yields a [Ne~V] bolometic correction of 8848.
Neither the [O~III] nor the [Ne~V] bolometric corrections require that the emission lines have been corrected for dust attenuation. In any case, we are unable to determine an accurate attenuation for the narrow line region with our current data due to large sky residuals near H$\beta$. We use the host galaxy A$_V$ value of 0.17 calculated in Section \ref{sec:stellar_continuum} to correct for attenuation in the narrow line region and find that [O~III] and [Ne~V] bolometric luminosities increase by a factor of 1.17, however this estimate is highly uncertain and thus we prefer to consider the uncorrected luminosities.

The bolometric luminosities computed from WISE W3 and W4 and the emission lines of [Ne~V] and [O~III] are reported in Table~\ref{lum_table}.  The 4 different tracers yield an average of 
$\log L_{bol} = 46.52 \pm0.46$, with the emission 
line luminosities being 0.1 - 0.2 dex lower than the IR luminosities.  This may be due to a small contribution of light from the companion source to the WISE bands.  J0048-0046 was included in a sample of IR-bright Dust Obscured galaxies  \cite[IR-bright DOGs;][]{Toba_2017} whose bolometric luminosities were estimated using SED fitting. They report log$_{10}$($L_{bol}$) $ = 46.26$ ergs s$^{-1}$, slightly lower than the values we find.
However, the overall agreement among the various tracers of bolometric luminosity is very good, considering the large amount of scatter seen in the galaxies used to calibrate bolometric corrections \citep[$\pm 0.3$~dex;][]{Duras_2020}.

Using the bolometric luminosity calculated from the [O~III] band, we compare to the QLF from \cite{Shen_2020} (cf. Figure 4 in their work) and find that J0048-0046 is $\sim$ 3 times more luminous than a typical quasar at $z=1$. We also note that J0048-0046 is slightly less luminous than the the median bolometric luminosity of core ERQs (log$_{10}$($L_{\text{bol}}$) $\sim 47.1$ ergs~s$^{-1}$) which were estimated using the W3 band.

We compute the radio luminosity of J0048-0046 using the FIRST reported flux density at a central observed-frame frequency of $\nu_{\text{obs}} = 1.435$ GHz. Following  \cite{Hwang_2018}, we calculate the radio luminosity as 

\begin{equation}
    \label{kcorrection}
    \nu L_{\nu} = 4\pi D_{\text{L}}^{2}(1+z)^{-1-\alpha}(\nu/\nu_{\text{obs}})^{1+\alpha}\nu_{\text{obs}} F_{\nu_{\text{obs}}}
\end{equation}

where the \textit{k}-correction is applied to the rest-frame frequency, $\nu = 2.77$ GHz, of J0048-0046 and a spectral index of $\alpha = -0.7$ is assumed. $F_{\nu_{\text{obs}}}$ is the observed flux density and $D_{\text{L}}$ is the luminosity distance calculated using our assumed cosmological parameters. 

The radio luminosity is reported in Table \ref{lum_table}. The threshold for an ERQ to be considered radio-loud is estimated by \cite{Hwang_2018} to be $\nu L_{\nu}[5 \text{ GHz}] > 10^{41.8} ~\text{erg~s}^{-1}$ given their average [OIII] luminosity \citep{Xu_1999}. This threshold criterion establishes J0048-0046 as a radio-quiet AGN. 

\begin{deluxetable}{ccccc} 

    \tablewidth{0.7\textwidth}
    \tablecaption{ J0048-0046 Optical, Infrared, and Radio Properties\label{lum_table}}
    \tablehead{\colhead{Name} & \colhead{Wavelength} & \colhead{log$_{10}$(L)} & \colhead{EW} & \colhead{log$_{10}$(L$_{bol}$)} \\
        & & \colhead{(erg s$^{-1}$)} & \AA & \colhead{(erg s$^{-1}$)} 
        }
    \startdata
        Mg~II & 2798 \AA &  $41.92_{-1.25}^{+0.31}$ & $19.2\pm17.1$ &- \\\relax
        [Ne~V] & 3345 \AA & $42.01_{-0.63}^{+0.34}$ & $22.7\pm27.9$&- \\\relax
        [Ne~V] & 3426 \AA & $42.44_{-0.42}^{+0.21}$& $62.9\pm39.2$& $46.39_{-0.42}^{+0.21}$ \\\relax
        [O~II] & 3728 \AA & $42.15_{-0.26}^{+0.16}$& $28.6\pm13.19$ &- \\\relax
        [Ne~III] & 3869 \AA & $42.08_{-0.45}^{+0.37}$ & $16.7\pm22.6$ &- \\\relax
        [O~III] & 5007 \AA & $43.17_{-0.24}^{+0.16}$ & $221.6\pm32.8$ & $46.43_{-0.30}^{+0.22}$ \\\relax
        H${\alpha}$ & 6562 \AA & $42.89_{-0.22}^{+0.15}$& $123.2\pm21.3$ & - \\
        \tableline
        W1 & 3.4 $\mu$m & $44.97_{-0.01}^{+0.01}$& \hyp{} & \hyp{} \\
        W2 & 4.6 $\mu$m & $45.26_{-0.01}^{+0.01}$ & \hyp{} &\hyp{} \\
        W3 & 12 $\mu$m & $45.66_{-0.02}^{+0.02}$ & \hyp{} & $46.59_{-0.06}^{+0.05}$ \\
        W4 & 22 $\mu$m & $45.93_{-0.03}^{+0.03}$ & \hyp{} & $46.65_{-0.07}^{0.05}$ \\
        \tableline
        FIRST & 1.435 GHz & $41.03_{-0.04}^{+0.04}$ & \hyp{} &\hyp{} \\
    \enddata
    \tabletypesize{\footnotesize}

    \tablecomments{Luminosities, emission line equivalent widths, and bolometric luminosities calculated from SDSS+NIRSPEC, WISE, and FIRST. Properties of the [O~III] doublet, [Ne~V], and H$\alpha$ are calculated using the combined best-fit profiles (see \S\ref{sec:emlines}).}
\end{deluxetable}

\section{Spectral Analysis} 
\label{sec:spectral_analysis}

\subsection{Emission Lines}
\label{sec:emlines}

\begin{figure}[h]
    \centering
    \includegraphics[width=\linewidth]{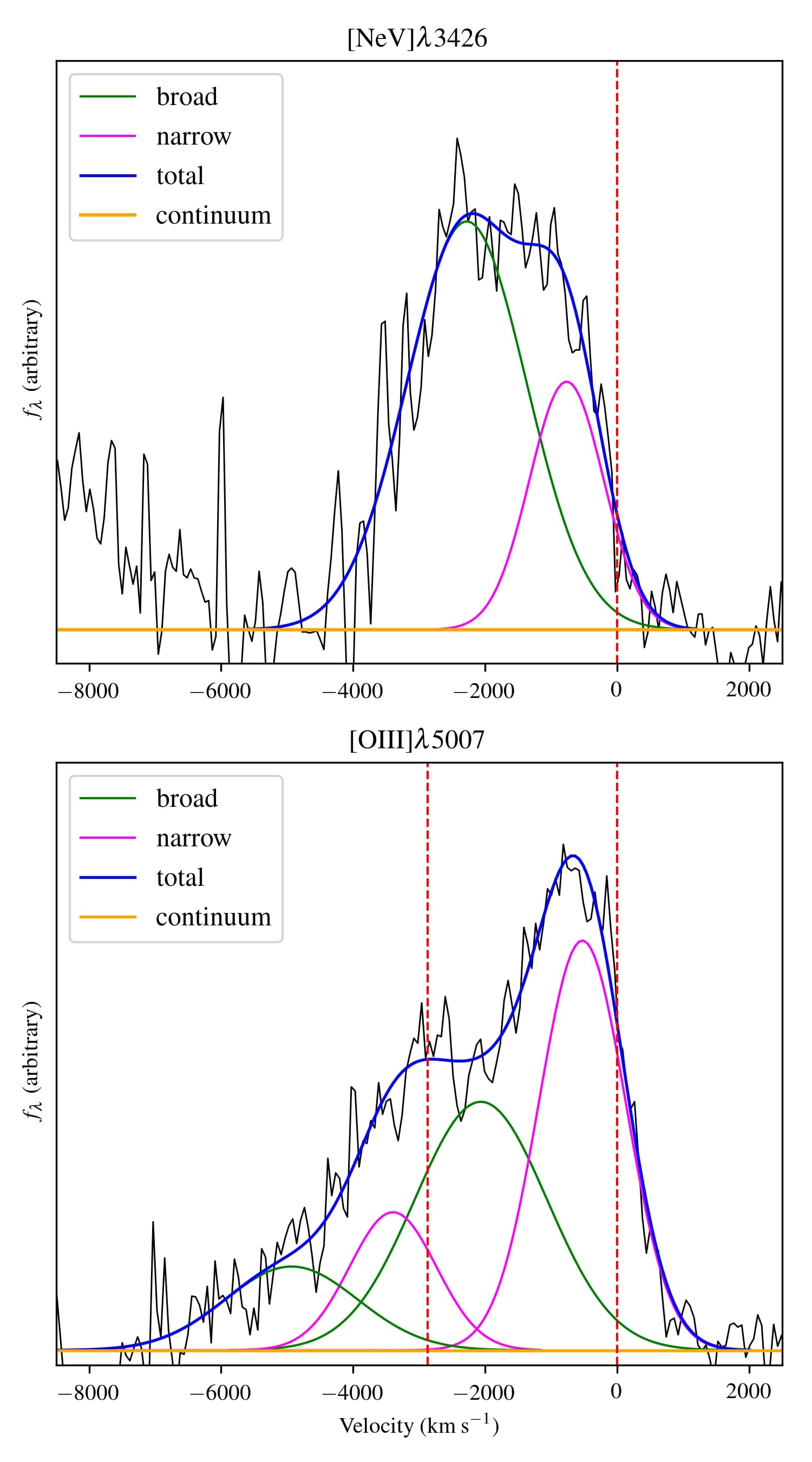}
    \caption{Velocity profile of [Ne~V] $\lambda3426$ from the SALT spectrum and [O~III] $\lambda5007$ from the BOSS spectrum. The red line indicates the emission lines' rest-frame zero velocity. The y-axis is in arbitrary units of flux since we are not able to flux calibrate with the SALT data.}
    \label{nev_line}
\end{figure}

We display the BOSS and NIRSPEC spectra jointly in Figure \ref{spectrum}. Upon initial visual inspection of the BOSS spectrum, it is apparent that the optical light of J0048-0046 is composed of two main components. The continuum light of J0048-0046 is dominated by starlight from the host galaxy (discussed in detail in  \S\ref{sec:stellar_continuum}). On top of this strong host galaxy continuum are emission lines typical of Type 2 AGN, specifically [Ne~III] $\lambda$3869, [Ne~V] $\lambda\lambda$3345, 3426, and [O~III] $\lambda\lambda$4959, 5007.
Due to sky-noise at 4861~\AA, we are not able to comment on the line profile of H$\beta$ in our SDSS spectrum and H$\alpha$ is heavily blended with [N~II].  However, we do see Mg~II $\lambda2798$, a line which is typically broad in Type~I AGNs. 
While Mg~II is highly broadened (FWHM$\sim$36\AA), it does not show the classical Lorentzian shape common in  Type I AGN, rather it appears to have a similar velocity structure to the lines associated with the NLR.  This is similar to core ERQs, where the UV broad emission lines (e.g, N~V, S~IV, C~IV) are found to lack broad wings. \citet{Hamann_2017} suggest that these unusual line kinematics are the result of outflows and spatially extended broad line regions that may connect to the low-density forbidden line regions farther out.


There are peculiarities present in the NLR of J0048-0046 that are apparent with an initial visual inspection of Figure \ref{spectrum}. One of these peculiarities is the strong blueshift and broadening in the NLR (Figures \ref{spectrum} and \ref{nev_line}). In fact, this causes [O~III]$\lambda$5007 to blend together with [O~III]$\lambda4959$. In contrast, the [O~II] $\lambda3728$ is narrow (FWHM $\sim7$ \AA) and shows no blueshift to its centroid. We use the multi-component fitting method described in \cite{Perrotta_2019} to decompose the line profile and calculate $w_{90}$ and $v_{98}$ of [O~III] and [Ne~V].  The $w_{90}$ parameter is a measure of the velocity interval containing 90\% of the line flux, excluding the most blue/red shifted regions of the line. The $v_{98}$ parameter is measured as 98\% of the line flux from positive velocities to negative velocities with respect to the systemic redshift. It can thus be thought of a near-maximum estimate of the outflow velocity. To estimate the error on $w_{90}$ and $v_{98}$ due to uncertainties in the fit, we consider the best-fitting parameters uncorrelated, vary them in a range of $\pm1\sigma$ and calculate the resulting change in $w_{90}$ and $v_{98}$. We adopt the maximal variation of these values as an upper limit error with typical value of 200–300 km~s$^{-1}$. An additional fit is performed on the H$\alpha+$[N II] emission by allowing a varying broad H$\alpha$ component while assuming a $1:3$ ratio between the two [N II] features, along with fixed-width narrow components set equal to that of [O II] ($7 \AA$), therefore giving us a cleaner estimate of the H${\alpha}$ luminosity and EW for Table \ref{lum_table}.


 The retrieved $w_{90}$ of [O~III] is $4008 \pm 378$ km~s$^{-1}$, very close to the median $w_{90}$ of 4050 km~s$^{-1}$ for the core ERQs. We retrieve an [O~III] $v_{98}$ of $-4095 \pm 423$~km~s$^{-1}$, slightly more blueshifted than the core ERQ median of $-3528$ km~s$^{-1}$. The line fitting for [Ne~V] yields a $w_{90}$ of $3469 \pm 300$ km~s$^{-1}$ and a $v_{98}$ of $-4013 \pm 300$ km~s$^{-1}$. Similarities in both parameters between the two lines suggests that they are likely part of the same outflow. The rest equivalent width (REW) of [O~III]$\lambda5007$ is 218\AA, approximately 100\AA\ higher than the median REW of the \cite{Perrotta_2019} sample, which is partially driven by the relatively faint continuum. 


\begin{figure}[ht]
    \centering
    \includegraphics[width = \linewidth]{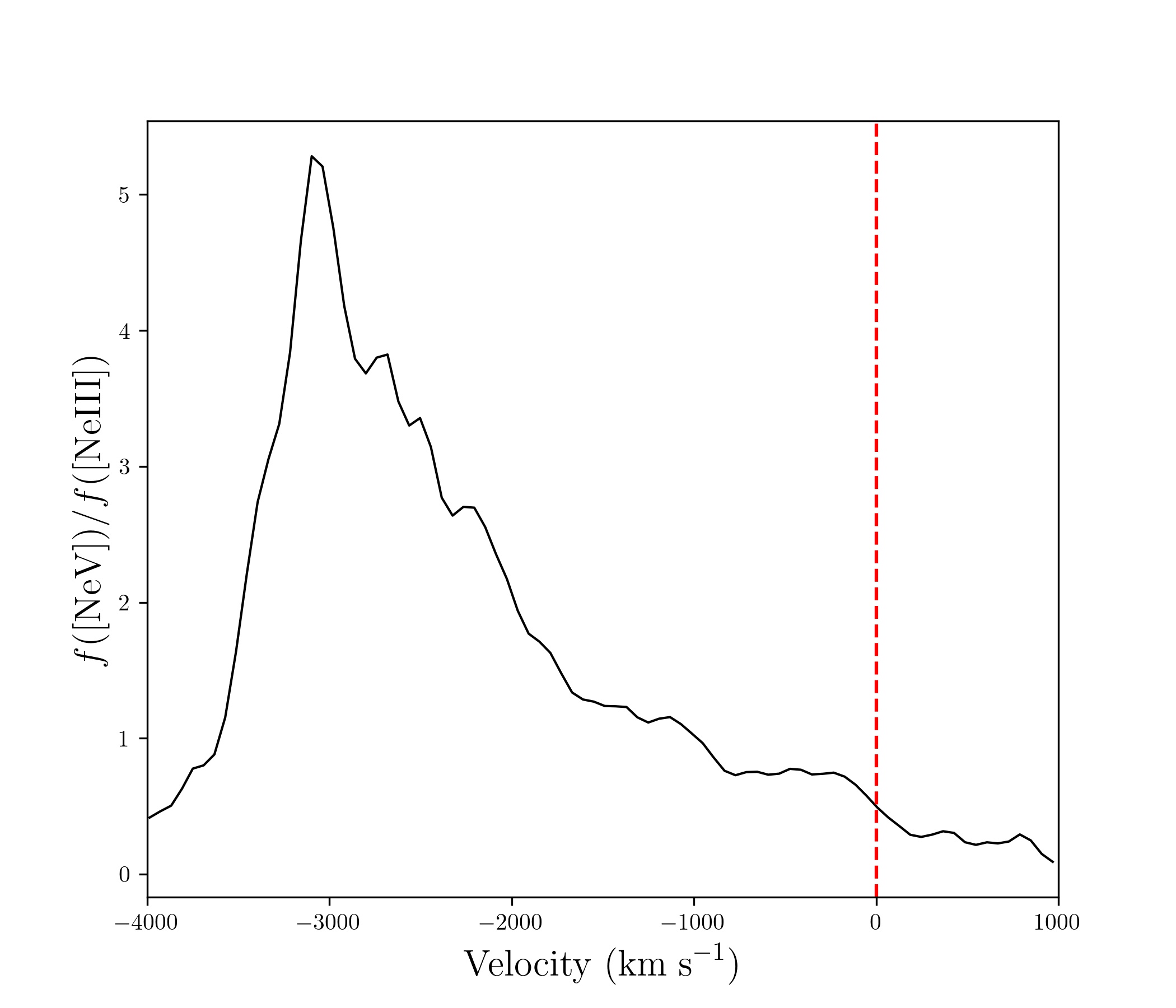}
    \caption{Continuum-subtracted flux ratio of [Ne~V]$\lambda3426$ to [Ne~III]]$\lambda3869$ from the BOSS spectrum smoothed by a ten pixel boxcar. Both emission lines have been transformed to velocity space based on their respective center wavelengths.}
    \label{fig:nev_neiii}
\end{figure}

Another indicator of quasar feedback observed in J0048-0046 is the atypical emission line ratios of the NLR. We report emission line luminosities in Table \ref{lum_table}. We correct for foreground Milky Way dust attenuation but do not correct for attenuation intrinsic to the source, as the S/N in the region of H${\beta}$ is too low. For comparison, we use the reported line ratios of the Type II quasar spectrum from \cite{Zakamska_2003}. The ratio of [O~III]$\lambda$5007/[O~II]$\lambda$3727 is 6.7 in J0048-0046 compared to 1.8 from the Type II composite. Additionally, the ratio of [Ne~V]$\lambda$3426/[Ne~III]]$\lambda$3869 is 2.95, whereas in the Type II composite it is 0.68. When considering the profile of the [Ne~V] to [Ne~III] ratio (Figure \ref{fig:nev_neiii}), we find that higher outflow velocities have higher [Ne~V]/[Ne~III], indicating a link between the the outflow and the ionization state of the gas.  [Ne~V] is an excellent indicator of AGN activity given its high ionization potential of 97~eV \citep{Gilli_2010}, however this line can also arise from photoionization by Wolf-Rayet stars found in young starbursts \citep{Abel_2008}.
Such starbursts have been observed to be young ($\sim3$ Myr) with extreme star-formation rate surface densities ($\Sigma_{\text{SFR}} \approx 3000 \text{~M$_{\odot}$~}\text{yr~}^{-1}\text{kpc}^{-1}$; \citealt{Sell_2014}). Given the many signs of AGN activity in J0048-0046 and a lack of a powerful starburst (see  \S\ref{sec:stellar_continuum}), we infer that [Ne~V] predominantly traces AGN activity in this source.

\subsection{Broadband Spectral Energy Distribution} 
\label{sec:sed_analysis}
\begin{figure*}[hbt!]
     \centering
     \includegraphics[width=\linewidth]{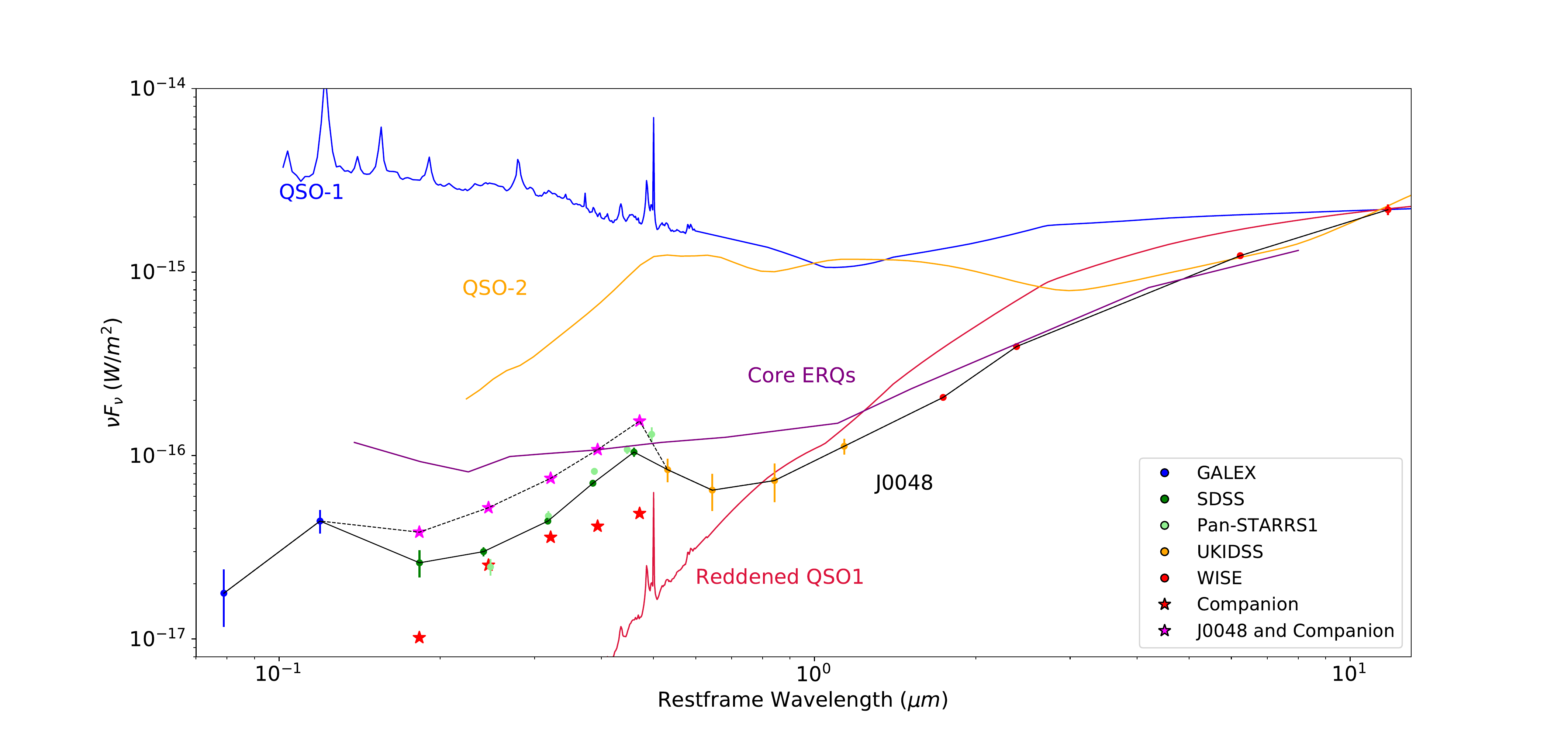}
     \caption{Restframe SED of J0048-0046 (black line, colored circles) compared to the core ERQ sample of \citet[][purple]{Hamann_2017}. Type 1 (blue) and Type 2 (yellow) quasar templates from \citet{Polletta_2007} and \citet{Hickox_2017} are shown for comparison, along with a reddened Type 1 quasar ($E(B-V) = 1$; red). SEDs are normalized to a common value at $10~\mu$m for visualization purposes. The star symbols show the optical SED of the companion galaxy (red) and its combined flux with J0048-0046 (magenta). The two galaxies are unresolved in the WISE and GALEX bands (see \S\ref{sec:sed_analysis}).  J0048-0046's SED is most similar to the SED of the core ERQs.  However, it is somewhat redder and it shows a bump at 0.45 $\mu$m due to the contribution of starlight to the SED.
     \label{fig:sed}}
 \end{figure*}

 Photometry over a wide baseline of wavelengths contribute to the SED of J0048-0046: GALEX, SDSS 
 Pan-STARRS1, UKIDSS, and WISE photometric measurements (\S\ref{sec:data}) build the quasar's spectral shape from the far UV to the IR. 
In Figure~\ref{fig:sed}, we compare J0048-0046's SED to the SEDs of several types of quasars. We use a Type 1 QSO composite template from \citet{Polletta_2007} and a Type 2 QSO template from \citet{Hickox_2017}.

 As noted in \S\ref{sec:data}, the apparent merging companion of J0048-0046 cannot be deblended in the WISE and GALEX data and likely contributes to the observed flux at these wavelengths. In Figure~\ref{fig:sed}, we show the fluxes of J0048-0046 and the companion in the SDSS bands as green circles and red stars, respectively;  the sum of the two fluxes is shown as magenta stars.   The companion's \textit{z} band flux ($\lambda_{rest}\sim4600$~\AA) is only 
 22\% that of J0048-0046 and it is undetected the UKIDSS YJHK bands. In addition, it is 0.5 mag bluer in \textit{g-i}  suggesting that it is lower mass and less dusty.  To quantify the companion's potential contribution to the WISE and GALEX bands, we use the \verb|Prospector| code \citep{leja19,johnson21} to fit the galaxy's SDSS photometry
 and to predict it's full SED.  The code models both the stellar light and dust reprocessed emission.  We use the same models and priors used in Davis et al. (submitted) and we assume the companion is at the same redshift as J0048-0046 on account of the tidal bridge between the two (Fig.~\ref{nev_image}). \verb|Prospector| employs gridless, bayesian parameter estimation with the incorporation of robust posterior sampling algorithms. 
 We randomly draw models from the probability distribution function and synthesize their GALEX and WISE magnitudes. Taking the 84th percentile of the simulated magnitude distribution, we compute the potential contribution of the companion to the flux in each band. The GALEX FUV samples wavelenghts below the Lyman break, so the contribution is negligible; in the NUV it is $<7$\%. In the WISE bands the companion contributes $4$\% in W1, $1$\% in W2, $0.5$\% in W3, and $0.1$\% in W4 to the total flux. Nonetheless, it may be most appropriate to consider the SED of the merging pair, which is shown as the dashed black line.
 
 Figure~\ref{fig:sed} shows that J0048-0046 has a dramatically different spectral shape than a typical Type 1 QSO and a Type 1 QSO behind a \cite{calzetti00} dust-reddening screen with $E(B-V) = 1$. There is a strong deviation in overall spectral shape which becomes particularly severe at wavelengths less than 6000~\AA.  This demonstrates that J0048-0046 is not consistent with being a typical Type~I quasar seen behind an unusually large dust reddening screen. 
 
The Type 2 QSO template is a reasonably good match to our source's SED at optical wavelengths on account of the significant starlight contribution in J0048-0046 (\S\ref{sec:stellar_continuum}). However, the Type~II template is relatively flat at wavelengths redder than 0.5~$\mu$m whereas the flux of J0048-0046 increases by nearly an order of magnitude.   The notable bump in our sources SED around 0.45 $\mu$m is a feature of the starlight (which can be seen more clearly in Figure~\ref{fig:ssp_fit}).  The [O~III] line emission contributes strongly to the Pan-STARRS $y$-band, but  emission lines impact the other optical+NIR filters by less than 20\%. The prominent 0.45~$\mu$m SED bump is evident even when the photometry is corrected for the contribution of emission lines. The importance of emission lines to the broadband SED may be greater in the UV with the bump near 0.12~$\mu m$ potentially arising from Ly$\alpha$ + N~V $\lambda 1240$.

J0048-0046's SED is most similar to the SED of the core ERQs of \cite{Hamann_2017}. In particular, the SED rises strongly above 1~$\mu$m, indicating a a heavily obscured AGN, but is surprisingly flat 
below 1~$\mu$m compared to the expectation of a simple `foreground screen' dust reddening. This unusual SED shape could arise due a clumpy dust distribution where some of the UV light is completely absorbed and re-radiated in the IR and some is transmitted along less attenuated sightlines. It is evident that our source is slightly more reddened than the core ERQ sample, however it is well within the scatter of the individual core ERQ SEDs as shown in Figure 16 of \cite{Hamann_2017}. The primary difference in the SEDs is the bump near 0.45 $\mu$m which suggests that J0048-0046's optical spectrum is dominated by starlight. It is possible that our source may be a Type~II ERQ (see \cite{Ross_2015} for some additional examples of Type~II ERQs.) 
In this case, some of the UV light might be due to dust scattered QSO continuum. Alternatively, it   
could be a Type~I ERQ where the QSO continuum that does penetrate a clumpy dust distribution is much more heavily reddened than the host galaxy, allowing the latter to dominate.

The rest-frame ultraviolet to mid-IR color of J0048-0046 is $i-W3=5.2$ mag (AB), meeting the threshold of $i-W3>4.6$ mag employed by \cite{Hamann_2017}. We calculate that this color would be $i-W3=6.5$ mag if J0048-0046 were at the median core ERQ redshift ($z=2.5$), consistent with J0048-0046's overall redder SED profile. 

\section{Host Galaxy Analysis}
\label{sec:host_galaxy_analysis}
\subsection{Morphology}
\label{morphology}
\begin{figure}[ht!]
    \centering
    \includegraphics[width=.8\linewidth]{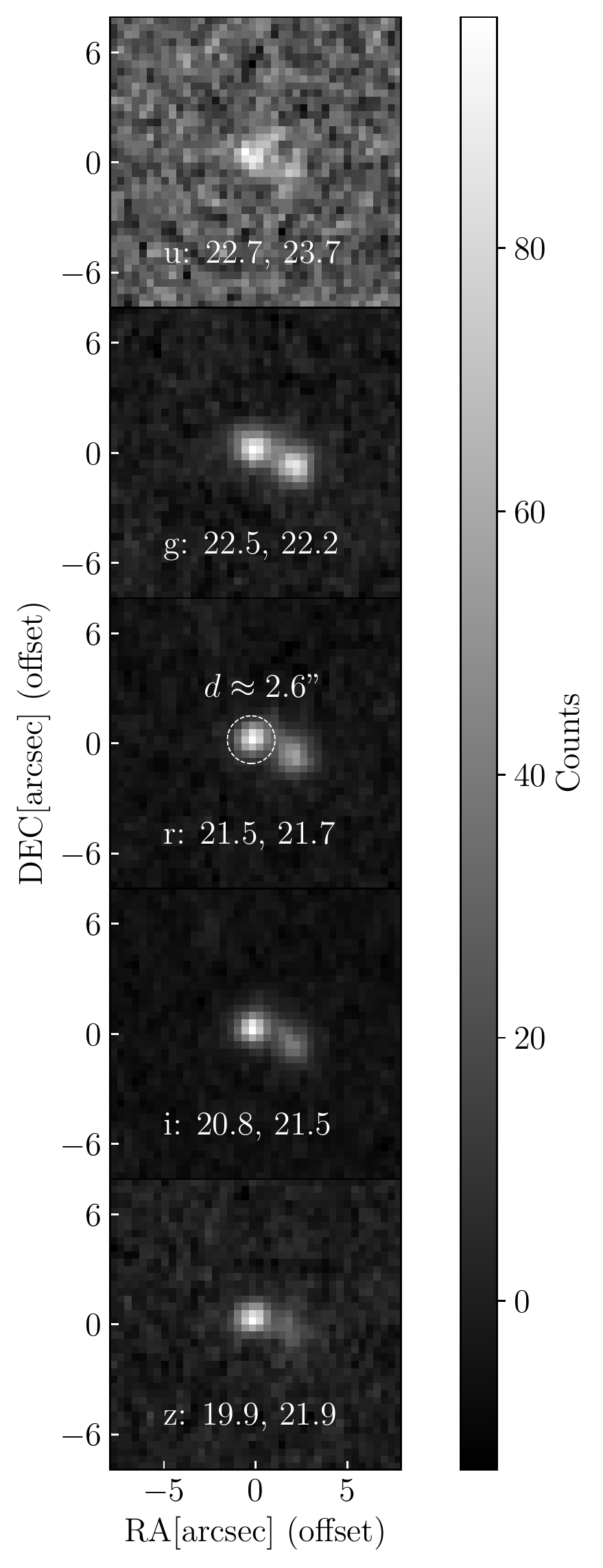}
    \caption{SDSS \textit{ugriz} images of J0048-0046 from the \cite{Fliri_2016} Stripe~82 co-addition. J0048-0046 is encircled in the \textit{r} band image. The fainter source to the southwest of J0048-0046 is a possible merging companion, as indicated by the apparent tail-like structure between the two galaxies. In each panel, numbers are the computed magnitudes from GALFITM for J0048-0046 (left) and the companion (right) (see Appendix \label{galfit}). The quoted magnitudes for J0048-0046 are the combination of its fitted sersic and psf magnitudes. The $ugriz$ filters probe restframe wavelengths of 1830, 2410, 3180, 3860, 4600~\AA\ respectively. 
    }
    \label{nev_image}
\end{figure}

J0048-0046's optical spectrum (Fig.~\ref{spectrum}) appears to be dominated by starlight (see \S\ref{sec:stellar_continuum} for a quantitative analysis).  This provides us with a unique opportunity to study the host galaxy of a candidate ERQ. 
In Figure \ref{nev_image} we show the deep \textit{ugriz} S82 co-add images \citep{Fliri_2016} of J0048-0046 and its neighboring galaxy. The median PSF FWHM is $\sim 1.1 "$ therefore we cannot resolve sub-structure smaller than $\sim$~8.7 kpc, limiting our analysis to very basic properties.

The most notable feature of the S82 co-add image, shown in Figure~\ref{nev_image}, is the existence of a lower luminosity source to the southwest of J0048-0046; the apparent bridge-like structure between the two indicates that it may be a merging companion.   However, we acknowledge the possibility that this apparent companion is simply a background source.  The companion galaxy is  2\farcs5 from J0048-0046, and was not included in the SDSS 2$\arcsec$ diameter fiber spectrum, so we have no information on its redshift at present. 


Resultant magnitude values from our GALFITM analysis of  the S82 co-add (\S\ref{sec:photo_data}) for both sources are reported in each panel of Figure \ref{nev_image}. We obtain a \textit{g} band $R_{50}$ value of $(0.30 \pm 0.05) "$ or $(2.42 \pm 0.40)$ kpc for J0048-0046 and $(0.52 \pm 0.05) "$ or $(4.20 \pm 0.1)$ kpc for the neighbor. The quoted errors are purely statistical \citep{Peng_2002} and do not reflect the uncertainty of our PSF model. The $R_{50}$ of J0048-0046 is roughly a factor of two lower than the average $R_{50}$ from the ten ERQs with HST-imaged host galaxies \citep{Zakamska_2019}.

\begin{figure}[ht!]
    \centering
    \includegraphics[width=\linewidth]{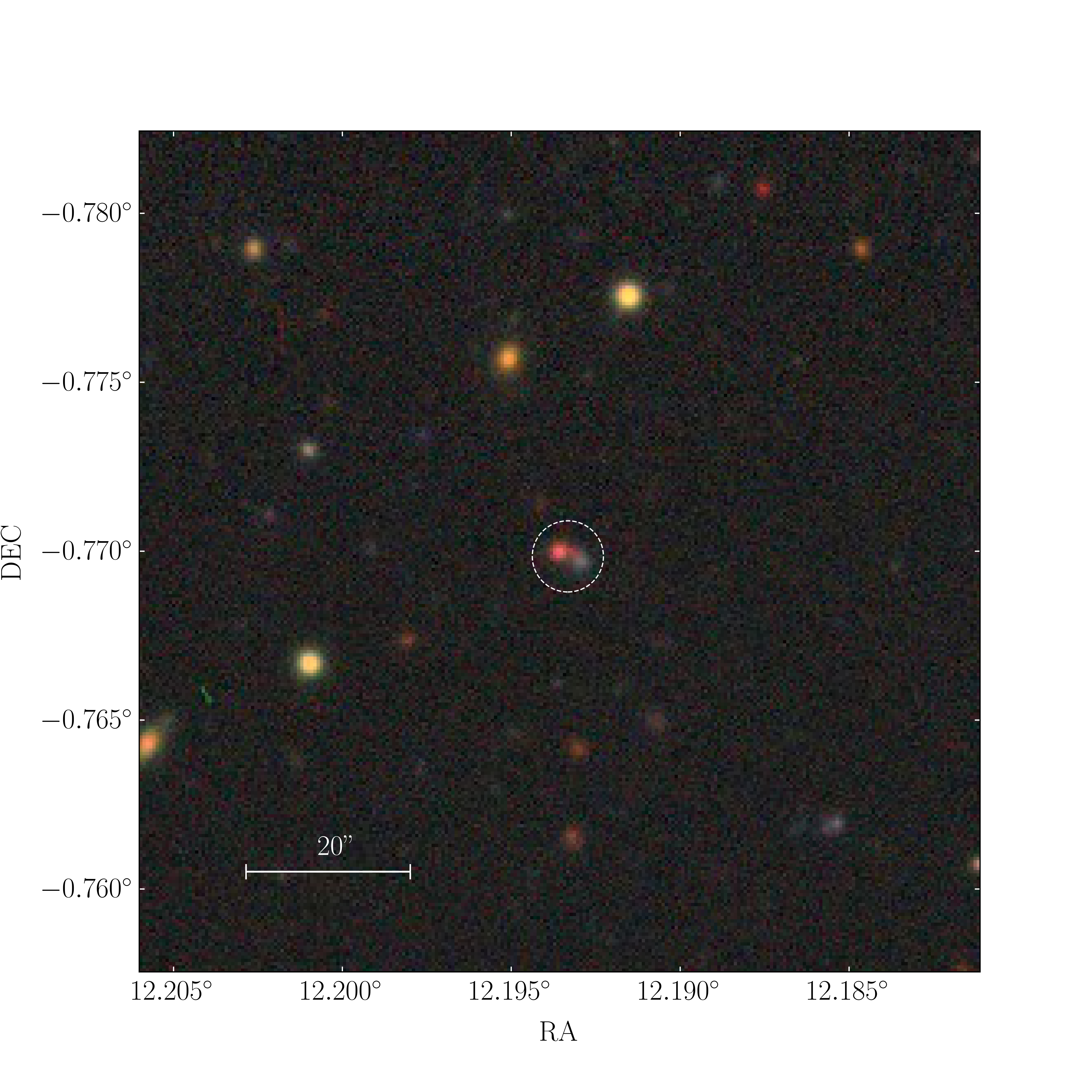}
    \caption{Color image from a combination of the DECaLS \textit{grz} images \citep{Dey_2019} showing the larger scale environment of  J0048-0046. J0048-0046 and its possible companion are encircled in the center.}
    \label{color_image}
\end{figure}

In Figure \ref{color_image}, we show a larger DECaLS color image centered on J0048-0046.  The companion appears significantly bluer than 
J0048-0046. Indeed, the SDSS ($g-i$) colors are 1.5 and 1.0 for J0048-0046 and the companion respectively.  The DECaLS image does not show any hints of a surrounding larger scale structure such as a galaxy cluster or massive group.

\subsection{Stellar Mass and Star Formation History} \label{sec:stellar_continuum}

The fact that J0048-0046 is dominated by host galaxy light presents us with a rare opportunity to investigate the stellar population of an ERQ host.   We fit the combined SDSS and NIRSPEC spectrum with stellar population synthesis models.  We use simple stellar population (SSP) models generated with the Flexible Stellar Population Synthesis
code \citep{Conroy_2009, Conroy_Gunn_2010} using  Padova 2008 isochrones, a \citet{Salpeter_1955} initial mass function, and a new theoretical stellar library `C3K' (Conroy et al., in prep) with a resolution of $R\sim10,000$. Because the S/N of our spectrum is low ($\sim 3$ per pixel), we utilize a fairly sparse model grid, consisting of 12 solar metallicity SSPs with ages of 1, 5, 25, 100, 200, 315, 400, 500, 630, 1000, 2000, 5600 Myr.   (At $z=0.94$ the Universe is roughly 6.1 Gyr old.)  We convolve our template spectra to SDSS resolution ($R\sim 2000$) and then further convolve them to the galaxy's velocity dispersion which we vary between 
$\sigma   = 100 - 200$ km~s$^{-1}$.

\begin{figure*}
    \centering
    \includegraphics[width=\textwidth]{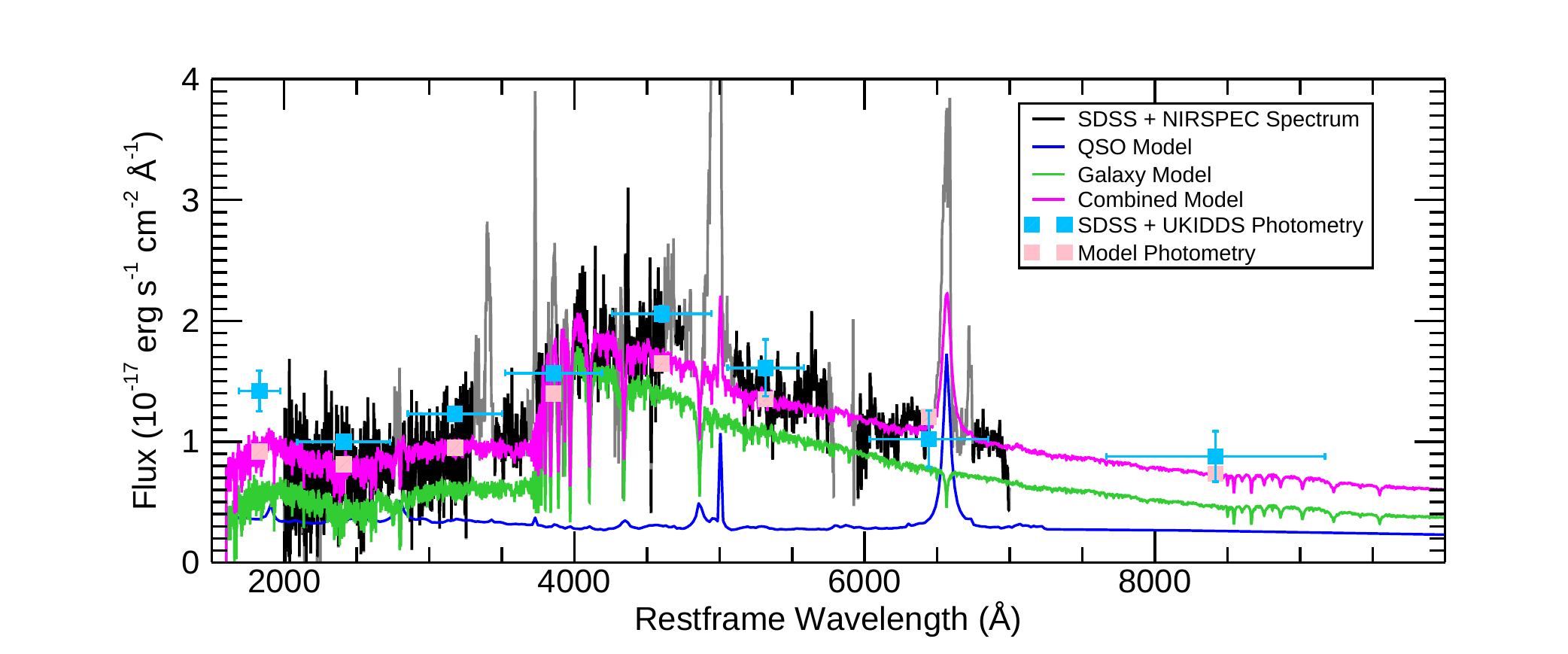}
    \includegraphics[width=\textwidth]{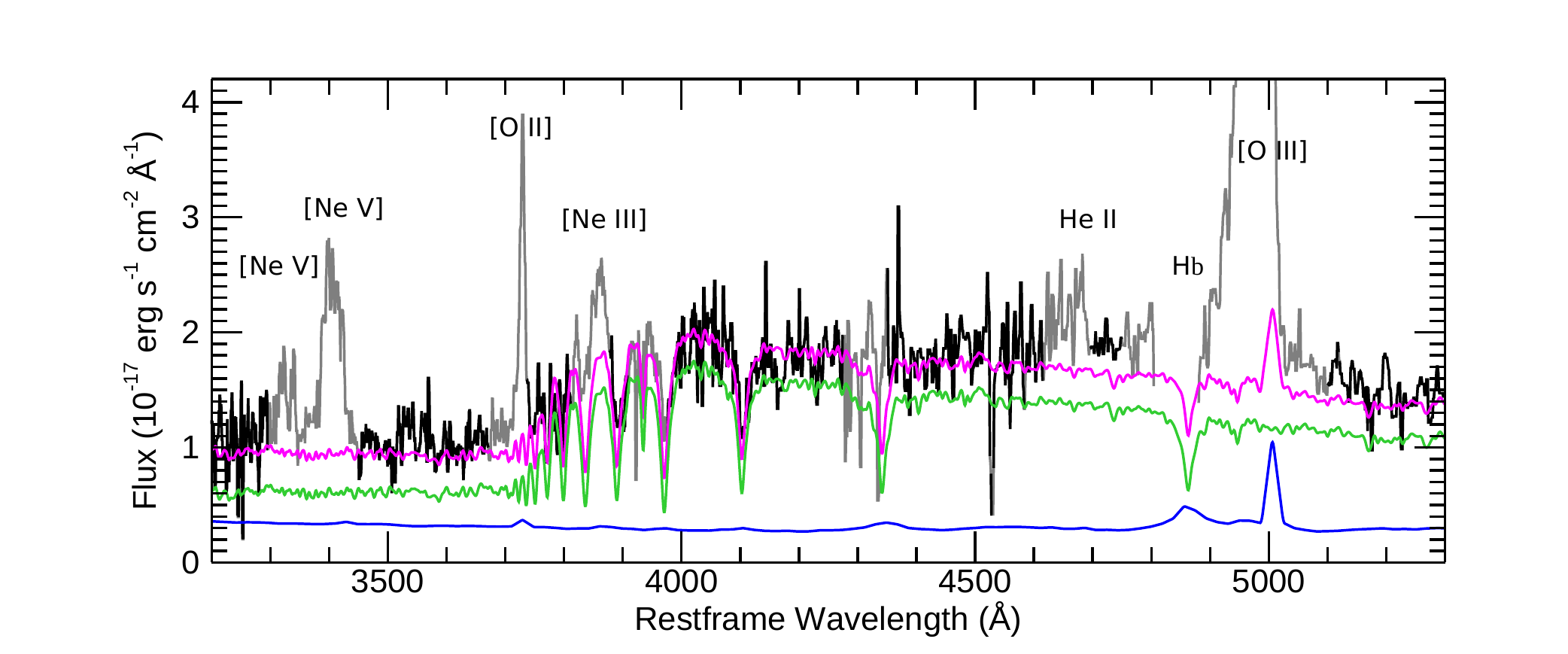}
    \caption{Combined SDSS and NIRSPEC spectrum (black) with the best fit stellar population synthesis model overlayed (magenta). Regions of the spectrum that are masked out of the fit are shown in grey.  The best fit model is made of a lightly attenuated stellar population (green) and a more heavily attenuated QSO spectrum (blue).  Photometric data from the SDSS Stripe 82 co-add ($ugriz$) and UKIDSS ($YJH$) are overlayed (blue points). We have normalized the spectrum to match the SDSS $i$-band photometry. Synthetic photometry of the best fit model is shown in pink.}
    \label{fig:ssp_fit}
\end{figure*}

Because our spectrum contains a mix of quasar and galaxy light, we also include a Type I quasar template from \citet{Polletta_2007}.  The quasar template was built from a combination of photometry and spectroscopy with the spectroscopic portion of the template extending to 6000 \AA.  We patch the template at wavelengths of 6000 - 7250~\AA\ with an SDSS quasar composite spectrum that we generated by selecting and stacking the spectra of high luminosity quasars ($\log L_{3000} > 45$) drawn from the \citet{Shen_2011} quasar catalog.   Since the H$\alpha$ line is masked out of our fit, the motivation for the patch is primarily aesthetic, when comparing the quasar and host galaxy light contributions (see Fig. \ref{fig:ssp_fit}).  

In our model, the stellar populations and the quasar experience different dust attenuation. For the stellar populations, we use a dust attenuation curve appropriate for massive high-$z$ galaxies from \citet{Salim_2018}. For the quasar, we use the AGN attenuation curve derived by \citet{Gaskell_Benker_2007} which lacks a 2200~\AA\ bump and is flatter than the curve we use for the host galaxy.  

After carefully masking regions of bad data and emission lines, we fit the model to our data using a custom built IDL code which is based on the \texttt{mpfitfun} routine \citep{Markwardt_2009}.  
The model returns the fraction of light contributed by each template (SSPs, QSO) at 5500 \AA\ and the reddening applied to the starlight and QSO light.  The fit is carried out over the full wavelength range of our SDSS + NIRSPEC spectrum, which is 2000 - 7000 \AA\ in the galaxy's restframe.  We considered including photometric data in our fit to extend the wavelength baseline, but our dust model is too simple (a foreground screen) and we do not model dust emission which would be required to reproduce the IR SED. To insure that the fit does not get stuck in a local minimum, we run our code 500 times with the starting parameters of the fit and the velocity dispersion randomly varied.  We then adopt the model with the lowest $\chi^2$ as our best fit. 

In Figure \ref{fig:ssp_fit},  we show our best fit models overplotted on the data. The QSO is highly reddened ($E(B-V) = 0.34,~A_V = 1.76$) while the starlight is minimally reddened ($E(B-V) = 0.06,~A_V = 0.17$). However, ERQs are known to have a large scattered light contribution in the UV \citep{Alexandroff_2018}, and thus the reddening computed for the QSO assuming our simple dust model should not be considered physically realistic. As we discussed in \S\ref{sec:sed_analysis}, a simple foreground screen dust model is unlikely to be able to reproduce the UV-IR SED of ERQs. The QSO contributes $\sim$16\% of the light at 4000 \AA, but more than 35\% below the Balmer break and at wavelengths longer than 8000 \AA.  While it is possible to obtain a good fit to the data without including the QSO template, this increases the reduced $\chi^2$ by 9\%.   


The stellar population is entirely dominated by a single SSP with an age of 400 Myr, suggesting the host galaxy is a `post-starburst' \citep[c.f.,][]{Wild_2009}.  Given that our code does not favor the inclusion of 300 or 500 Myr old templates in the fit, we can conclude that the burst had a short duration.  If we assume that the stellar mass ascribed to the 400 Myr old population formed at a constant rate over 100 Myr, it implies a star formation rate of 560 M$_{\odot}$~yr$^{-1}$, consistent with values found for sub-mm galaxies \citep[c.f.,][]{Dudzevivciute_2020}. 

Whether star formation is truly shut off at present is difficult to precisely ascertain from SSP fitting because there is some degeneracy between the QSO spectrum and a very young stellar population (e.g, both are relatively featureless and blue over the range of wavelengths considered here). An alternative approach to estimating the SFR in the last 10~Myr is to use the nebular emission lines. 
In J0048-0046, the H$\alpha$ line (Figs.~\ref{spectrum}, \ref{fig:ssp_fit}) is broad, blueshifted, and blended with  [N~II] $\lambda\lambda6548,6584$; its ionization is likely dominated by the AGN. However, [O~II]$\lambda\lambda3726,3729$ is relatively narrow  ($\text{FWHM}=7~\text{\AA}$) and appears to trace the host galaxy kinematics, suggesting that it does not arise in the AGN outflow. Unfortunately, [O~II] is far from an ideal SFR indicator due to its strong sensitivity to dust and metallicity \citep{Moustakas_2006}.  To partially account for this, \citet{Moustakas_2006} derive calibrations between SFR and $L_{\rm[OII]}$ (uncorrected for dust) in bins of B-band luminosity.  J0048-0046's host galaxy has $\log (L_B/L_{B,\sun}) = 11.09$, placing it in the highest luminosity bin defined by \citet{Moustakas_2006}. Using their SFR calibration and our measured $\log L_{[OII]}$ (Table~\ref{lum_table}), we find SFR$ = 72^{+45}_{-31}$ M$_{\odot}$yr$^{-1}$, where the error bars reflect the SFR calibration uncertainty.  
This should be considered an upper limit on the on-going SFR, as the [O~II] may partly arise from the AGN or from ionization by evolved stars \citep{Yan_2006}. Taking the derived numbers for the past and present value of the SFR at face value, we see that star formation has decreased by a factor of 5 - 14 over the past 400 Myr, even if star formation is still on-going at some level.  This confirms the post-starburst nature of the host galaxy.   The Lick H$\delta$ absorption index is 7.2 +/- 2.7,  consistent with the H$\delta  > 5$ criterion adopted by \citet{Alatalo_2016} for their sample of shocked post-starburts."   
 
The spectral fitting algorithm finds a M/L ratio which suggests a total stellar mass of $5.6 \times 10^{10}$~M$_{\odot}$ assuming a \citet{Salpeter_1955} IMF, however, this is likely a lower limit.   The fact that our code fails to find evidence of an older stellar population is, in part, due to a tendency of our code to use a minimum number of templates to fit the data. (A similar behavior is present in other stellar population fitting software such as pPXF \citep{Cappellari_2017}.) If we force our fit to put no more than 50\% of the mass at stellar ages $\leq 2$~Gyr, we increase the reduced $\chi^2$ by only 1\% since the 5.6 Gyr SSP contributes minimally (0.1-10\%) to the total stellar light.  With our young/intermediate age population thus limited, the stellar mass increases by a factor of 1.8 to $9.8 \times 10^{10}$~M$_{\odot}$, which we take to be our upper limit on the mass.   Plugging our plausible range of stellar masses ($5.6-9.8 \times 10^{10}$~M$_{\odot}$) into the \cite{Zahid_2016} scaling relation between stellar mass and velocity dispersion, we predict a velocity dispersion of 161 - 189~km~s$^{-1}$, whereas our measured dispersion is 140~km~s$^{-1}$.   However, the starlight in our galaxy is clearly dominated by the young burst, which is likely to be centrally concentrated and more rotation dominated \citep{Mihos_1994, Hopkins_2008b, Hopkins_2009a} and thus the measured dispersion may not trace the galaxy's full potential well.

\section{Variability} \label{sec:variability}
Upon an initial inspection of J0048-0046's optical spectrum, its prevalent narrow-line emission makes it tempting to classify the source as a Type 2 quasar. Through AGN unification models, it is expected (and observed) that Type 2 quasars show little to no optical variability due to obstruction around their coronal region \citep{Bershady_1998, Sarajedini_2006, Yip_2009}. This idea has persisted through a more recent analysis of Extreme Variability Quasars \citep[EVQs;][]{Barth_2014} with $\Delta g \approx 1$, which found that EVQs in Type 2 samples are actually Type 1.8 and 1.9 contaminants. Given that J0048-0046 is not a standard Type 2 quasar (\S\ref{sec:sed_analysis}), stellar population modeling favors a QSO contribution to its optical spectrum (\S\ref{sec:stellar_continuum}), and it is conveniently located in the S82 region (\S\ref{sec:photo_data}), we are motivated to investigate the source for variability. Moreover, no ERQ to date has been observed with high cadence, time-series photometry, making J0048-0046 the first ERQ to be investigated for variable properties.

Using the SDSS Data Release 16 (DR16) and NEOWISE data products described in Section \ref{sec:photo_data}, we construct light curves to inspect for variability in J0048-0046. Note that the SDSS \textit{u} band is dropped for variability inspection due to its large error bars. An initial examination of the light curve shows that J0048-0046 has strong, short time-scale variability in the optical regime. To quantify the variability across different photometric bands, we make use of the fractional variance \citep[$F_{\text{var}}$;][]{Vaughan_2003}, which calculates the root mean square (rms) variability amplitude as a percentage. When applied to the 11 epochs of DR16 photometry, we find $F_{\text{var}}$ values of $35\pm5\%$ in the \textit{g} band down to $29\pm4\%$ in the \textit{z} band.

\begin{figure} 
    \includegraphics[width=\linewidth]{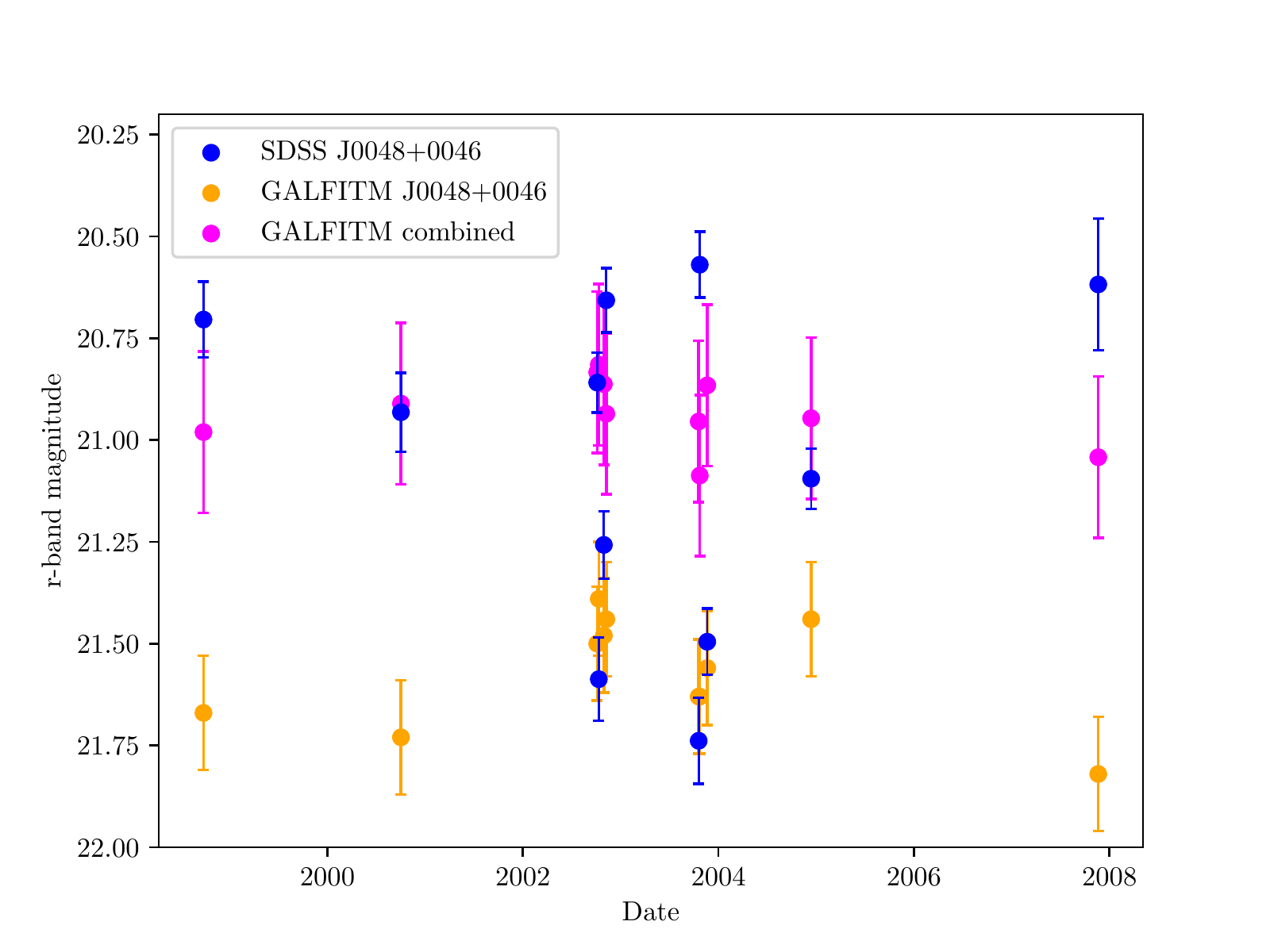}
    \caption{SDSS \textit{r} band lightcurve of J0048-0046 with blue points showing the model magnitudes taken directly from SDSS DR16 photometric catalog and orange points showing the output of GALFITM applied to the DR16 images (\S\ref{sec:photo_data}). The magenta points show the combined magnitudes of J0048-0046 and its companion as calculated by GALFITM. Variations in the deblending of the two sources by the SDSS photometric pipeline lead to apparent variability in the catalog data. With our more careful treatment of the data using GALFITM, no convincing evidence of variability is found.}
    \label{galfitLC}
\end{figure}

However, one concern with the SDSS catalog photometry is that sources are modeled and deblended separately at each epoch and differences in the deblending of J0048-0046 and its close companion could masquerade as variability.  To explore this, we carry out our own photometry of the 11 DR16 images using GALFITM as described in \S\ref{sec:photo_data}.  Figure \ref{galfitLC} shows the original J0048-0046 \textit{r} band light curve as queried from SDSS DR16 along with the light curve constructed using GALFITM. The running median of the GALFITM light curve is shifted by $\sim+0.7$ mag when compared to the SDSS DR16 catalog photometry and signs of variability are greatly reduced. 
Figure \ref{galfitLC} also shows the combined light curve of J0048-0046 and its companion; at certain epochs, the combined flux of the two sources is proximal to the DR16 catalog values for J0048-0046, indicating that pipeline deblending issues likely occurred. Since the variation of our GALFITM photometry is less than the computed errors, $F_{\text{var}}$ is undefined in each of the four bands meaning that there is no variability detected at restframe $0.24 - 0.46$~$\mu$m in our relatively shallow data.
Since ERQ samples might have high merger rates, any future ground-based photometric studies of ERQs should exercise similar caution when using survey data products. We find that variability is also undetectable for the NEOWISE W1 and W2 lightcurves ($N=149$) which probe restframe wavelengths of 1.7 and 2.4 $\mu$m respectively.


\section{Results and Discussion}
\label{sec:results}

Our analysis of the multi-wavelength photometry and optical spectra of J0048-0046 conclusively indicates that the source is not a typical Type 1 or 2 quasar, nor is it a Type 1 quasar behind a strong dust-reddening screen. Broad and blueshifted narrow lines ([O~III], [Ne~V]) with velocities as high as $-4000$~km~s$^{-1}$ suggest that the source's AGN may be powering galactic-scale outflows.  In these respects, the J0048-0046 resembles the core ERQ's studied by \cite{Hamann_2017}, however it is at a lower redshift such that it is easier to study its host galaxy.  Below we summarize the defining properties of core ERQs and their likely physical interpretation.  We then revisit the key aspects of J0048-0046 that lead us to classify it as an ERQ and discuss notable differences.  We highlight the unique aspects of our analysis -- the study of the QSOs variability and the host galaxy's stellar population -- and we consider how the insights gained from J0048-0046 can shed further light on the nature of ERQs. 


\subsection{The Nature of Core ERQs}

The core ERQ sample defined by \cite{Hamann_2017} is selected to have $i-W3 > 4.6$, $z = 2.0 - 3.4$, and  C~IV~EW $> 100$~\AA.  The resulting sample of core ERQs have the following properties:
\begin{itemize}
    \item unusual emission line flux ratios, in particular, large ratios of NV/Ly$\alpha$, N~V/C~IV and Si~IV/C~IV

    \item high EW broad emission lines with unusual wing-less line profiles and blueshifted line centroids

    \item SEDs that are red in the  restframe optical-IR but flat/blue in the UV and are, thus, inconsistent with quasars  behind a simple dust reddening screen 

    \item ubiquitous powerful ionized outflows traced by [O~III]~5007 \citep{Perrotta_2019} 

    \item low luminosity radio emission possibly associated with wind-driven shocks \citep{Hwang_2018} 

    \item high Eddington ratios \citep[$L_{bol}/L_{Edd} = 0.22 - 4.03$;][] {Perrotta_2019} 
    
    \item luminous host galaxies \citep[log($L_B/L_{B,\sun}) = 10.92 - 12.03$;][]{Zakamska_2019}
    
\end{itemize}

\citet{Hamann_2017} carefully consider several different models to explain these properties, in particular, the close connection observed between reddening and the unusual emission line properties. They converge on a model where ERQs are luminous quasars with unusually powerful and extended outflows that encompass most of the line-forming regions.  
In this model, the unusual UV emission line properties are a consequence of outflow-dominated broad line regions that are vertically extended above the accretion disc and potentially connected to the high-speed low-density [O~III] gas much farther out. The unusual SEDs are explained by patchy obscuration due to small dusty clouds embedded in the outflows. 
Spectropolarimetric observations of a small sample of ERQs support this general scenario \citep{Alexandroff_2018}. 
While this outflow model emphasizes small-scale phenomena and does not explicitly link the quasars to the evolution of their hosts, it has been suggested that ERQs could represent an transition phase in quasar/host galaxy evolution in which a heavily obscured quasar is blowing out the dusty circum-nuclear ISM and quenching star formation prior to becoming an optically luminous quasar. 

Alternatively, \cite{Villar_2020} suggest that the unusual spectral properties in core ERQs can be explained by their high bolometric luminosities and an intermediate orientation that results in the continuum emitting region and inner BLR being predominantly obscured whereas the outskirts of the BLR are visible. Hence, they suggest that the observed suppression in the UV and continuum along with the FWHMs of the broadlines (intermediate between Type 1 and 2 as found by \cite{Hamann_2017}) are explainable as an orientation effect. However, this scenario has difficulty explaining the broad blueshifted [O~III] lines that are ubiquitous in ERQs \citep{Perrotta_2019} but less common in normal luminous quasars. Notably, \cite{Perrotta_2019} estimate that the kinematic power of strong ERQ outflows outflows based on [O~III] emission are 1--1.5 dex higher than those of blue quasars with comparable luminosities.

Thus, there remains some ambiguity as to whether ERQs can be explained by geometry and orientation effects or whether they represent a particular outflow-dominated phase of quasar evolution.   If the latter, a key question is whether their strong outflows can regulate galactic processes in their hosts. 

In what follows, we summarize the physical properties of J0048-0046 and compare them to those of core ERQs. Most importantly, we consider the implications of our results on quasar-galaxy evolutionary scenarios. 



\subsection{J0048-0046's Extreme NLR Outflows}
\label{sec:extreme_outflows}

J0048-0046 meets the core ERQ color selection criterion with $i-W3=5.2$ mag (AB), however, it is at $z=0.94$ and therefore our optical+NIR spectrum does not cover C~IV~$\lambda1549$. In ordinary Type~I quasars, Mg~II $\lambda 2800$ and C~IV have broadly consistent properties \citep{vanden_berk_2001} albeit the C~IV line typically has a relative blueshift to the Mg~II line of a few hundred km s$^{-1}$ \citep{Richards_2002}. The Mg~II line in our spectrum has low S/N and a low EW (19~\AA) relative to the criterion defined for core ERQs (EW C~IV $>$ 100~\AA).   
However its ionization potential is much lower than that of C~IV (7.6 vs 47.9 eV, respectively), and ERQs are known to have stronger high ionization lines \citep{Hamann_2017}.  Nonetheless, the kinematics of the Mg~II line are very similar to those of C~IV in ERQs: the line is moderately broad (FWHM$\sim36$\AA) but lacks the broad wings typical of Type I quasars.  While the S/N of  Mg~II is low, it appears to have very similar kinematics to the lines associated with the NLR ([Ne~V], [O~III]). The H$\alpha$ line has a high EW (123 \AA), however the fact that these lines originate in both the BLR and NRL which make their kinematics more difficult to interpret. The H$\beta$ line is marred by sky residuals.    

The most striking feature of J0048-0046's spectrum is its strong broad [Ne~V]$\lambda\lambda3345, 3426$ emission ([Ne~V]$\lambda3426$ EW = 63~\AA).  
Unusually strong [Ne~V] emission has previously been found in some Type 2 quasars \citep{Rose_2015a, Yuan_2016}. \cite{Rose_2015a} deem such objects `coronal line forest forest AGNs.' These AGN additionally show signatures of [Fe~VII] $\lambda\lambda5721,6087$ which has a similar ionization potential to [Ne~V]. The latter transition is present, albeit weakly, in the NIRSPEC spectrum of J0048-0046 and shows a blueshift similar to [Ne~V]. 
\cite{Rose_2015a} suggest that these emission lines originate along the inner wall of the obscuring torus, which could support the orientation hypothesis discussed above.  However, the [Ne~V]-strong Type~II's previously studied have very narrow lines.  The extreme broadening ($\sim4000$~km~s$^{-1}$)  and blueshifting ($\sim-4000$~km~s$^{-1}$) of the [Ne~V] and [O~III] lines in J0048-0046 (Fig~\ref{nev_line}), suggest a different origin for the high ionization emission, perhaps in fast wind-shocks as suggested by \citet{Hwang_2018}.  

Lower redshift ERQs ($0.3 < z < 1.5$) with visibly identifiable [Ne~V] are present in 17 out of 41 AGN from the \cite{Ross_2015} sample where [Ne~V] falls within the wavelength coverage of BOSS. For 14 of these 17 AGN we are able to perform the multi-component fitting \citep{Perrotta_2019} on [Ne~V] as previously carried out in \S\ref{sec:spectral_analysis}. The median $w_{90}$ value of this sub-sample is $3057$ km~s$^{-1}$ with a standard deviation of $1303$ km~s$^{-1}$. The median $v_{98}$ value of this sub-sample is $-1738$ km~s$^{-1}$ with a standard deviation of $924$ km~s$^{-1}$. The [Ne~V] profile in J0048-0046 is then comparable to this sub-sample of ERQs in terms of broadening ($w_{90} = 3469 \text{~km~s}^{-1}$), however it exhibits the strongest outflow ($v_{98} = -4013 \text{~km~s}^{-1}$). The [O~III] EW is also very large (221~\AA) in J0048-0046, comparable to the values found in the core ERQ sample studied by \citet{Perrotta_2019}. The [O~III] and [Ne~V] kinematics are very well matched overall (Fig~\ref{nev_line}). Notably, the higher velocity gas has a higher [Ne~V]/[Ne~III] ratio (Fig.~\ref{fig:nev_neiii}), suggesting a link between the outflow velocity and its distance from the ionization 
source. \cite{Perrotta_2019} find that larger [O~III] line widths correlate with redder colors which suggests that dust may play a role in driving ERQ winds (see their Fig.~6). The [O~III] $w_{90}$ and $i-W3$ color of J0048-0046 place it squarely on this correlation.


Overall, we conclude that while the C~IV EW of J0048-0046 is unknown, the strength of the high ionization lines and the similar [O~III] kinematics to the core ERQs studied in \citet{Perrotta_2019} make it likely that J0048-0046 is a low redshift core ERQ. The observed correlation between ionization and outflow velocity adds support to the hypothesis that the unique emission line properties are linked to the energetics and geometry of the outflows. 


\subsection{J0048-0046's SED}
\label{sec:sed_color}


The SED of J0048-0046 is in stark contrast to those of typical Type 1 QSOs and highly reddened Type 1 QSOs. The optical-NIR portion is in strong agreement with Type 2 QSOs due to its strong starlight component, however it has a substantially larger MIR flux. Overall, the SED of J0048-0046 best matches the composite SED of core ERQs from \cite{Hamann_2017}, although it slightly redder. The strong stellar contribution to the SED and lack of variability at NUV-NIR wavelengths (\S\ref{sec:variability}) indicates that J0048-0046 is either a Type~II ERQ or a Type~I where the reddened QSO continuum contributes minimally at these wavelengths.


In \S\ref{sec:luminosities} we noted that J0048-0046 was included in a sample of IR-bright Dust-Obscured galaxies \citep[DOGs][]{Toba_2017}. These were selected by a color of $i-W4>7\text{ mag}$ in the redshift range of $0.05 < z < 1.02$ with detected [O~III]$\lambda5007$ in their BOSS spectra. Of the 36 sources studied in their analysis, J0048-0046 exhibits the second-most powerful outflow in [O~III] and is 4$\sigma$ higher than the mean of this sample. Their utilized color selection is similar to ERQ color selections and they note that 4 of their 36 AGN overlap with the \cite{Ross_2015} ERQs. They find that IR-bright DOGs have similar, unusually large ionized outflows in [O~III]. However, the IR-bright DOGs have a median bolometric luminosity of 46.02 erg~s$^{-1}$, roughly one percent that of the core ERQs. One possibility is that ERQs and IR-bright DOGs are part of an evolutionary sequence with DOGs representing a somewhat more obscured and less active earlier phase, with J0048-0046 near the border line between the two.

\subsection{Radio Emission as a Byproduct of Shocks}
\label{sec:radio_emission}
    We found from the FIRST radio flux measurement of J0048-0046 (Section \ref{sec:luminosities}) that the quasar is radio quiet. The origin of radio-quiet emission in quasars is still not fully understood as radio emission in the radio-quiet regime is too low to be explained by powerful jets originating from the AGN as observed in some quasars. 
    A multitude of phenomena have been shown to contribute to radio-quiet emission in quasar populations, such as radio synchrotron emission as a byproduct of star formation feedback \citep{Helou_1985, Fan_2016, Farrah_2017} young compact jets found in the host \citep{Fanti_1990}, or accelerated particles produced from shocks as a result of AGN-driven winds \citep{Zakamska_2014, Nims_2015}. 
    Multiple scenarios for radio quiet emission in core ERQs were considered by \cite{Hwang_2018} and they concluded that AGN-driven winds are the most plausible explanation for the observed radio luminosities. In considering star formation as the origin of radio-quiet emission, they indicate that an SFR of 2700 M$_{\odot}$yr$^{-1}$ \citep{Bell_2003} would be required to produce the radio luminosities of ERQs. The resulting rest-frame U-band luminosities of 10 core ERQ hosts from \cite{Zakamska_2019} suggest that ERQs are not undergoing such high SFRs. Our SFR upper limit of 72 M$_{\odot}$yr$^{-1}$ (\S\ref{sec:stellar_continuum}) further indicates that star formation is likely not the dominant cause of radio quiet emission in ERQs. 


    \cite{Hwang_2018} further demonstrated that core ERQs extend the relationship between [O~III] line-width and radio luminosity originally shown for the $z < 0.8$ type 2 quasar sample in \cite{Zakamska_2014}, suggesting that AGN-driven winds in ERQs can reach into the surrounding host galaxies and shock the ISM. Using our calculated radio luminosity (Table~\ref{lum_table}) from the reported FIRST flux density and the measured $w_{90}$ of [O~III] (4062 km s$^{-1}$), we find that J0048-0046 lies in the same region as ERQs in Figure 7 of \cite{Hwang_2018}.
   
   \subsection{J0048-0046's Host Galaxy}
\label{sec:stellar_properties}
     To understand the quasar--host galaxy interaction in core ERQs,  it is critical to measure the physical conditions in the host galaxy. 
     However, this has proven to be difficult due to the redshift of the core ERQs and their dominant nuclear point sources. \emph{HST} imaging of ten $z=2-3$ ERQs detected only six host galaxies well enough to measure very basic structural parameters, and the absence of deep multi-color imaging precluded any assessment of the stellar population properties \citep{Zakamska_2019}.

     While our stellar population fitting exercise is limited by the S/N of the spectrum, it is aided by the fact that the QSO contributes only 15 - 35\% of the light at restframe wavelengths of 2000-7000 \AA. 
     We find that the host galaxy experienced a strong starburst $\sim400$~Myrs ago which was subsequently quenched.  Assuming that ERQs represent very young AGNs, the age of the starburst implies a delay in the onset of AGN activity relative to star formation. This is consistent with the scenario proposed by
      \citet{Wild_2010} where the black hole accretion rate peaks $\sim$250~Myr after a merger-induced starburst. 
      They postulate that the black hole is fueled by 
      mass lost by intermediate mass stars in the bulge, but accretion of this material is suppressed at early times by supernova feedback.
      
     
     The original sample of 65 ERQs from \cite{Ross_2015} includes seven quasars ($0.34 < z < 1.095$) which show signatures of ongoing/recent starburst phases, three of which are post-starbursts based on their strong Balmer breaks. Notably, these three quasars are included in the broad/blue-shifted [Ne~V] comparison sample discussed in \S\ref{sec:extreme_outflows}. Given that post-starbursts make up just $\sim2$\% of the galaxy population at these redshifts \citep{Wild_2016}, this fraction is quite remarkable. Similar to J0048-0046, these sources may be useful cosmic laboratories to study the relationship between powerful quasar winds and the impact they have on their hosts. 
     
      The host galaxy of J0048-0046 is luminous, with log($L_{B}/L_{\odot}$) = 11.1. This is slightly lower than the median (log($L_{B}/L_{\odot}$) = 11.2) but well within the range of the quasar-subtracted host luminosities of the ten HST-imaged core ERQs studied in \cite{Zakamska_2019}. These $z\sim 2.5$ core ERQs have host galaxy luminosities of 4$L_{B}^{*}$, whereas J0048-0046's host is roughly 5$L_{B}^{*}$, using $L^{*}_{B}$ at $z\sim1$ from  \citet{Giallongo_2005}. In contrast, the six Type 2 quasars studied by \cite{Zakamska_2019} have host luminosities around $L^{*}$.
     
      We estimate that the host galaxy of J0048-0046 has a stellar mass between $5.6-9.8 \times 10^{10}$~M$_{\odot}$. This mass is within the range ($10^{10.4} - 10^{11.4}$ M$_{\odot}$) found for core ERQs by \citet{Zakamska_2019}. Plugging our mass into the empirical relation between black hole mass and bulge mass found for Type I quasars at z=0.2 - 1 \citep{Ishino_2020}, we infer a black hole mass between  $2.5 - 4.8 \times 10^{8}$~M$_{\odot}$.  Notably, the Ishino et al. $M_{*}-M_{BH}$ relationship displays a large scatter (0.35 dex) and predicts a lower black hole mass than scaling relations derived for local quiescent galaxies \citep{Kormendy_Ho_2013}. This scatter, combined with the fact that J0048-0046 has a large uncertainty for its stellar mass, an unknown bulge to disk ratio, and the fact that the Ishino et al. scaling relation is derived from measurements of quasars which are unambiquosly of Type I, means that our estimate of the black hole mass is likely on the order of a magnitude uncertain. Computing the Eddington luminosity for our inferred black hole masses yields $\log L_{Edd} = 46.5 - 46.8$.  Comparing these values to the bolometric luminosities given in Table \ref{lum_table} suggests that the quasar is radiating near its Eddington luminosity with $L_{bol}$/$L_{edd}\sim1.23$, consistent with an average $L_{bol}$/$L_{edd}=1.03$ in the \cite{Perrotta_2019} core ERQs.
     
     Another interesting result found by \cite{Zakamska_2019} was that only 20-30\% of their HST imaged ERQs showed merger signatures, in contrast to the high fraction expected at $z=2-3$. They suggest that the near-Eddington accretion and strong feedback of ERQs are associated with relatively late stages of mergers which have early-type remnants.  However, they note that merger signatures could be hidden at optical wavelengths or that gas instabilities within ERQ disks could be an alternate triggering mechanism. The apparent tidal bridge feature between J0048-0046 and its companion (Fig.~\ref{nev_image}) indicates that a merger event may have been the contributing factor to its star-formation history and its eventual high AGN accretion rate.  The overall picture for ERQs will become more clear as we build up larger samples of ERQs with high resolution imaging. 
     
     
     
     A key result of our SSP analysis was our finding of low dust attenuation in J0048-0046's host galaxy ($A_V$=0.17). 
     This indicates that the heavy extinction/reddening in core ERQs may be representative solely of dust-obscuration around the AGN and it disfavors the proposed evolutionary scenario \citep{Glikman_2007, Glikman_2012, Glikman_2013, Urrutia_2009, Banerji_2012, Assef_2013} where the ERQs are transitioning between a dusty sub-mm galaxy and a luminous QSO. The post-starburst nature of the host galaxy likewise contradicts this simple evolutionary paradigm.
     
    An interesting question is whether the QSO may be helping to maintain the galaxy in its post-starburst state. Using our [O~III] luminosity and $v_{98}$ value, we follow the exact procedure laid out in Section 5.1 of \cite{Perrotta_2019} to estimate the outflow's kinetic power in J0048-0046. We obtain log($\dot{E_{k}}$) $ = 45.4$ erg~s$^{-1}$, within the range of the \cite{Perrotta_2019} core ERQ sample and shifted above other quasar samples (see \cite{Perrotta_2019}, Figure 9). This yields $\dot{E_{k}}/L_{bol} = 6-9\%$ depending on the adopted bolometric luminosity from Table \ref{lum_table}. 
    Various simulations have suggested that values at or above five percent of the total radiated energy are required for the outflow to have a significant impact on the host and quench/regulate its star formation \citep{Scannapieco_2004, Di_matteo_2005, Hopkins_2005, Hopkins_2008b}. Hence, the powerful outflow J0048-0046 exhibits has the potential to greatly disrupt the ISM of the host and may have a connection to its post-starburst state.
     In order to further consider the relationship between galactic-scale properties and quasar energetics, future stellar population analysis of ERQs host galaxies, including the sub-sample from \cite{Ross_2015} mentioned above, are necessary to consider their host's evolutionary state in relation to their AGN-powered winds. 
     
     \section{Summary \& Conclusions}
\label{sec:summary}

Apparent peculiarities in the optical spectrum and the near-UV to far-IR SED of the $z = 0.94$ quasar, J0048-0046, present it as a distinct source in the context of typical Type 1 and 2 quasars. With a more detailed analysis (\S\ref{sec:spectral_analysis}), we find that it is most likely an Extremely Red Quasar \citep[ERQ;][]{Ross_2015, Hamann_2017}, a candidate young quasar that has just recently begun AGN activity and is in the process of blowing out gas at extreme velocities. While properties of these recently discovered ERQs have been studied and discussed extensively \citep{Zakamska_2019, Perrotta_2019}, J0048-0046 has the key advantage of having a spectroscopically resolvable host-galaxy to constrain its star-formation history in addition to high-cadence, multi-epoch photometry to investigate for variability. Both of these features make J0048-0046 unique in the context of ERQs and can offer insight into the nature of young AGN and their properties.

J0048-0046 is a luminous AGN ($L_{bol} = 10^{46.5}$ erg~s$^{-1}$) that exhibits powerful outflows in the NLR [O~III]$\lambda5007$ and [Ne~V]$\lambda3426$ with velocity widths (w90) around 4000 km $^{-1}$ (see \S\ref{sec:spectral_analysis}, Fig.~\ref{nev_line}). The [O~III] velocity width is inline with those displayed in core ERQs \citep{Perrotta_2019}. It shows unusually high line ratios of [Ne~V]$\lambda$3426/[Ne~III]$\lambda$3869 and [O~III]$\lambda$5007/[O~II]$\lambda$3728 compared to typical quasars. Comparing the line profiles of [Ne~V] and [Ne III], we find that higher ionization occurs at larger outflow speeds which indicates a connection between the outflow and the ionizing process. Additionally, its radio-quiet emission could possibly arise from wind-shocks driven by the powerful outflow (\S\ref{sec:radio_emission}). The connection between high ionization lines with high velocities supports the idea that the peculiar emission line properties in ERQs are linked to outflows rather than orientation. 

J0048-0046 meets the optical to mid-IR color selection criterion of $i - \text{W3} > 4.6$ imposed to select ERQs in \cite{Hamann_2017}. An analysis of the source's SED 
(\S\ref{sec:sed_analysis}, Fig.~\ref{fig:sed}) clearly shows that it is different than both standard and heavily reddened Type 1 quasars. Its dominance by starlight in the optical to NIR gives it a similar shape to Type 2 quasars at these wavelengths, but its SED deviates strongly in the MIR. We find that J0048-0046 is most similar to the median SED of the ``core'' ERQs from \cite{Hamann_2017}, albeit slightly redder. 
    
In \S\ref{sec:stellar_continuum} we take advantage of the strong host galaxy continuum apparent in J0048-0046 (Figure \ref{fig:ssp_fit}) to characterize the stellar population of an ERQ for the first time. We find that the massive host galaxy ($\log (M/M_{\sun}) = 10.7-11$) underwent a strong starburst 400 Myr ago before it was abruptly quenched. If ERQs are indeed young AGN, then this timescale could indicate a connected offset between the starburst and the beginning of AGN activity as suggested by \citet{Wild_2010}. 
The low extinction in the host is in contrast to expectations from galactic evolutionary models and may suggest that ERQs are not dusty sub-mm galaxies in transition.

Based on J0048-0046 and the neighboring galaxy's morphology, it is possible that the post-starburst phase of J0048-0046 has arisen from a major merger. Given the rareness of post-starbursts \citep[$\sim2\%$ of all massive galaxies at $z\sim1$;][]{Wild_2016} and ERQs \citep[$<1\%$ of luminous AGN;][]{Hamann_2017}, it seems unlikely that J0048-0046 is a member of both populations by coincidence. Instead, our estimated time-lag between peak starburst and AGN states (again, inline with \citet{Wild_2010}) makes an evolutionary connection between ERQs and post-starbursts seem more likely. Future SSP analysis of larger ERQ samples with resolvable host galaxies are necessary to solidify these findings.

The multi-epoch SDSS data release and S82 catalog photometry initially present J0048-0046 as a variable AGN. However, careful photometric analysis using GALFITM indicates that successful deblending of J0048+0046 and its companion are not always produced by the SDSS pipeline. We find that J0048-0046 is not variable at restframe 0.24 - 2.4 $\mu$m wavelengths. Combined with the dominance of host galaxy light in the optical spectrum, this may point to J0048-0046 being a Type 2 ERQ or a Type I where the contribution of the AGN continuum to the total light at these wavelengths is minimal.

We conclude that J0048-0046 is a $z \sim 1$ core ERQ and poses a unique opportunity to study the stellar populations the first time in such an object. The strongly broad and blueshifted [Ne~V] is an indicator that the AGN is driving strong, high ionization outflows. In future work, we will attempt to spatially resolve this outflow and constrain the extent to which it reaches into the host galaxy of J0048-0046 \citep{Tremonti_2020}.

\acknowledgments
The authors thank the anonymous referee for constructive comments which greatly improved the paper. 
This material is based upon work supported by the National Science Foundation (NSF) under a collaborative grant (AST-1814233, 1813299, 1813365, 1814159 and 1813702). R.~C.~H. acknowledges support from the NSF through CAREER award no. 1554584. S.~P.\ and A.~L.~C.\ acknowledge funding by the Heising-Simons Foundation grant 2019-1659.

Funding for SDSS-III has been provided by the Alfred P. Sloan Foundation, the Participating Institutions, the National Science Foundation, and the U.S. Department of Energy Office of Science. The SDSS-III web site is http://www.sdss3.org/. SDSS-III is managed by the Astrophysical Research Consortium for the Participating Institutions of the SDSS-III Collaboration including the University of Arizona, the Brazilian Participation Group, Brookhaven National Laboratory, Carnegie Mellon University, University of Florida, the French Participation Group, the German Participation Group, Harvard University, the Instituto de Astrofisica de Canarias, the Michigan State/Notre Dame/JINA Participation Group, Johns Hopkins University, Lawrence Berkeley National Laboratory, Max Planck Institute for Astrophysics, Max Planck Institute for Extraterrestrial Physics, New Mexico State University, New York University, Ohio State University, Pennsylvania State University, University of Portsmouth, Princeton University, the Spanish Participation Group, University of Tokyo, University of Utah, Vanderbilt University, University of Virginia, University of Washington, and Yale University.

\software{Astropy \citep{2013A&A...558A..33A}, PyRAF \citep{pyraf2012}, REDSPEC \citep{kim2015}, MPFIT \citep{Markwardt_2009}, FSPS \citep{Conroy_2009, Conroy_Gunn_2010}, Prospector code \citep{leja19,johnson21}, GALFIT \citep{Peng_2002}, GALFITM \citep{Vika_2013}}

\begin{appendix}
\section{GALFITM parameters for SDSS co-add}
    \begin{deluxetable}{cccccccc}[!ht] 
    
        \tablewidth{0.72\textwidth}
        \tablecaption{GALFITM Parameters \label{galfit}}
        \tablehead{\colhead{Filter} & \colhead{Object} & \colhead{Model} & \colhead{Flux} & \colhead{$R_{\text{e}}$} & \colhead{Sersic Index} & \colhead{$b/a$} & \colhead{P.A.} \\
        \colhead{} & \colhead{} & \colhead{} & \colhead{mag} & \colhead{''} & \colhead{} & \colhead{} & \colhead{deg}
        }  
         
        \startdata 
            \textit{u} & J0048 & psf & $23.40\pm0.05$ & - & - & - & - \\
            \textit{g} & J0048 & psf & $23.28\pm0.05$ & - & - & - & - \\
            \textit{r} & J0048 & psf & $22.32\pm0.05$ & - & - & - & - \\
            \textit{i} & J0048 & psf & $21.54\pm0.05$ & - & - & - & - \\
            \textit{z} & J0048 & psf & $20.70\pm0.05$ & - & - & - & - \\
            \tableline
            \textit{u} & J0048 & sersic & $23.60\pm0.05$ & $0.07\pm0.10$ & $1.69\pm2.52$ & $0.91\pm0.39$ & $58.55\pm30.59$ \\
            \textit{g} & J0048 & sersic & $23.23\pm0.05$ & $0.30\pm0.05$ & $1.52\pm1.33$ & $0.62\pm0.20$ & $42.35\pm16.03$ \\
            \textit{r} & J0048 & sersic & $22.27\pm0.05$ & $0.66\pm0.04$ & $1.29\pm0.47$ & $0.40\pm0.06$ & $25.25\pm5.22$ \\
            \textit{i} & J0048 & sersic & $21.49\pm0.05$ & $1.11\pm0.06$ & $1.06\pm0.32$ & $0.31\pm0.03$ & $11.15\pm2.52$ \\
            \textit{z} & J0048 & sersic & $20.65\pm0.05$ & $1.71\pm0.1$ & $0.79\pm0.31$ & $0.32\pm0.05$ & $-1.65\pm3.46$ \\
            \tableline
            \textit{u} & neighbor & sersic & $23.70\pm0.04$ & $0.51\pm0.06$ & $3.05\pm1.03$ & $0.14\pm0.18$ & $-57.09\pm9.51$ \\
            \textit{g} & neighbor & sersic & $22.24\pm0.04$ & $0.52\pm0.02$ & $1.92\pm0.43$ & $0.58\pm0.05$ & $-49.08\pm5.05$ \\
            \textit{r} & neighbor & sersic & $21.71\pm0.04$ & $0.57\pm0.03$ & $0.94\pm0.22$ & $0.78\pm0.05$ & $-48.01\pm4.97$ \\
            \textit{i} & neighbor & sersic & $21.50\pm0.04$ & $0.59\pm0.03$ & $0.36\pm0.28$ & $0.65\pm0.05$ & $-55.59\pm4.81$ \\
            \textit{z} & neighbor & sersic & $21.88\pm0.04$ & $0.61\pm0.10$ & $0.14\pm0.91$ & $0.15\pm0.21$ & $-73.25\pm7.62$ \\
        \enddata
        \tabletypesize{\footnotesize}
    
        \tablecomments{Best-fit GALFITM parameters for J0048 and its neighboring galaxy, applied to the S82 coadd \textit{griz} images. Column }
    \end{deluxetable}
\end{appendix}

\end{document}